\documentclass[fleqn,usenatbib]{mnras}

\usepackage{fix-cm}
\usepackage{newtxtext,newtxmath}

\usepackage[T1]{fontenc}

\DeclareRobustCommand{\VAN}[3]{#2}
\let\VANthebibliography\thebibliography
\def\thebibliography{\DeclareRobustCommand{\VAN}[3]{##3}\VANthebibliography}

\usepackage{graphicx}	
\usepackage{amsmath}	
\usepackage[table]{xcolor}
\usepackage{upgreek}
\usepackage{float}
\usepackage{longtable}
\usepackage{caption}
\usepackage{xspace}
\usepackage{booktabs}
\usepackage{siunitx}
\usepackage{multirow}

\newcommand{\rone}{FRB~20121102A\xspace}

\newcommand{\rsixseven}{FRB~20201124A\xspace}
\newcommand{\roneoneseven}{FRB~20220912A\xspace}
\newcommand{\meightone}{FRB~20200120E\xspace}
\newcommand{\ronefourseven}{FRB~20240114A\xspace}

\newcommand{\sigproc}{{\tt SIGPROC}\xspace}
\newcommand{\presto}{{\tt PRESTO}\xspace}

\definecolor{agreen}{cmyk}{0.80,0.20,0.80,0.00}

\newenvironment{Contfigure*}{%
\addtocounter{figure}{-1}%
\begin{figure*}}{%
\end{figure*}}

\def\torun{Toru\'n\xspace}
\def\nancay{Nan\c{c}ay\xspace}
\newcommand{\dmunit}{pc\,cm$^{-3}$\xspace}
\newcommand{\rom}[1]{\uppercase\expandafter{\romannumeral #1\relax}}

\makeatother
\usepackage[switch]{lineno} 

\newcommand{\ASTRON}{ASTRON, Netherlands Institute for Radio Astronomy, Oude Hoogeveensedijk 4, 7991 PD Dwingeloo, The Netherlands}
\newcommand{\CUT}{Department of Space, Earth and Environment, Chalmers University of Technology, Onsala Space Observatory, 439 92, Onsala, Sweden}
\newcommand{\MITK}{MIT Kavli Institute for Astrophysics and Space Research, Massachusetts Institute of Technology, 77 Massachusetts Ave, Cambridge, MA 02139, USA}
\newcommand{\UVA}{Anton Pannekoek Institute for Astronomy, University of Amsterdam, Science Park 904, 1098 XH Amsterdam, The Netherlands}
\newcommand{\TORUNI}{Institute of Astronomy, Faculty of Physics, Astronomy and Informatics, Nicolaus Copernicus University, Grudziadzka 5, 87-100 \torun, Poland}
\newcommand{\STKI}{Astropeiler Stockert e.V., Astropeiler 1-4, 53902 Bad M\"{u}nstereifel, Germany}
\newcommand{\OXFORD}{Astrophysics, The University of Oxford, Keble Road, Oxford, OX1 3RH, UK}
\newcommand{\TSI}{Trottier Space Institute, McGill University, 3550 rue University, Montr\'eal, QC H3A~2A7, Canada}
\newcommand{\MCGILL}{Department of Physics, McGill University, 3600 rue University, Montr\'eal, QC H3A~2T8, Canada}
\newcommand{\GRUNN}{Kapteyn Astronomical Institute, University of Groningen, Kapteynborg 5419, 9747 AD, Groningen, The Netherlands}
\newcommand{\SRON}{SRON Netherlands Institute for Space Research, Niels Bohrlaan 4, 2333CA Leiden, The Netherlands}

\title[Energetics of \roneoneseven]{A probe of the maximum energetics of fast radio bursts through a prolific repeating source}

\author[Ould-Boukattine et al.]{O.~S.~Ould-Boukattine,$^{1, 2}$\thanks{E-mail: ouldboukattine@astron.nl}
P.~Chawla,$^{1,2}$
J.~W.~T.~Hessels,$^{2,1,3,4}$
A.~J.~Cooper,$^{5}$
M.~P.~Gawro\'nski,$^{6}$
\newauthor W.~Herrmann,$^{7}$
D.~M.~Hewitt,$^{2}$
J.~Huang,$^{3,4}$
D.~Huppenkothen,$^{2,8}$
F.~Kirsten,$^{9,1}$
D.~C.~Konijn,$^{10,1}$
\newauthor K.~Nimmo,$^{11}$
Z.~Pleunis,$^{2,1}$
W.~Puchalska$^{6}$
and M.~P.~Snelders$^{1,2}$
\\
\\
$^{1}$\ASTRON\\
$^{2}$\UVA\\
$^{3}$\TSI\\
$^{4}$\MCGILL\\
$^{5}$\OXFORD\\
$^{6}$\TORUNI\\
$^{7}$\STKI\\
$^{8}$\SRON\\
$^{9}$\CUT\\
$^{10}$\GRUNN\\
$^{11}$\MITK
}

\date{Accepted 2025 October 25. Received 2025 October 13; in original form 2025 July 30 }

\pubyear{2025}

\begin{document}
\label{firstpage}
\pagerange{\pageref{firstpage}--\pageref{lastpage}}
\maketitle

\begin{abstract}
Fast radio bursts (FRBs) are sufficiently energetic to be detectable from luminosity distances up to at least seven billion parsecs (redshift $z > 1$). Probing the maximum energies and luminosities of FRBs constrains their emission mechanism and cosmological population. Here we investigate the maximum energetics of a highly active repeater, \roneoneseven, using 1,500\,h of observations. We detect $130$ high-energy bursts and find a break in the burst energy distribution, with a flattening of the power-law slope at higher energy -- consistent with the behaviour of another highly active repeater, \rsixseven. There is a roughly equal split of integrated burst energy between the low- and high-energy regimes. Furthermore, we model the rate of the highest-energy bursts and find a turnover at a characteristic spectral energy density of $E^{\textrm{char}}_{\nu} = 2.09^{+3.78}_{-1.04}\times10^{32}$\,erg\,Hz$^{-1}$. This characteristic maximum energy agrees well with observations of apparently one-off FRBs, suggesting a common physical mechanism for their emission. The extreme burst energies push radiation and source models to their limit: at this burst rate a typical magnetar ($B = 10^{15}$\,G) would deplete the energy stored in its magnetosphere in $\sim$ 2150\,h, assuming a radio efficiency $\epsilon_\mathrm{radio} = 10^{-5}$. We find that the high-energy bursts ($E_\nu > 3 \times 10^{30}$\,erg\,Hz$^{-1}$) play an important role in exhausting the energy budget of the source.
\end{abstract}




\section{Introduction}

Fast radio bursts \citep[FRBs;][]{petroff_2022_aarv} are observed with durations ranging from microseconds \citep{snelders_2023_natas} to seconds \citep{chime_2022_natur} and fluences ranging from about 0.01\,Jy\,ms to 1000\,Jy\,ms \citep{spitler_2016_natur, kirsten_2024_natas}. Of the thousands of known FRBs, most are observed as one-off events and only a few percent are known to repeat \citep{chime_2023_apj}. It remains unclear whether the repeaters and apparent non-repeaters have different astrophysical origins. An exceptionally bright (MJy) burst has been observed from the Galactic magnetar SGR~1935+2154, and strongly suggests that at least some FRBs originate from magnetars \citep{chime_2020_natur_galacticfrb, bochenek_2020_natur}. 

Though FRB emission is almost certainly beamed, the beaming fraction is unknown. For simplicity, isotropic-equivalent energies are used to compare different FRBs even though this is an overestimation of the total energy released in radio waves. Given the typical Mpc- to Gpc-distances of FRBs \citep{gordon_2023_apj}, their isotropic-equivalent energies are known to range from roughly $E = 10^{37} - 10^{42}$\,erg. This is at least four orders-of-magnitude larger than what is measured for Galactic radio pulsars, including the giant pulses seen from the Crab pulsar \citep{bera_2019_mnras}. SGR~1935+2154, however, is known to produce bursts whose energies span the range between pulsars and FRBs (Ref.~ \citep{kirsten_2021_natas} and references therein).

Furthermore, it is expected that the detected radio energy is only a small fraction ($\sim10^{-5}$) of the total bolometric energy released at the source during an FRB event \citep{metzger_2019_mnras,mereghetti_2020_apjl}. Taking this into account, the most energetic known FRBs may be associated with $\sim10^{46-47}$\,erg events that are comparable to the most extreme `giant flares' from magnetars \citep{kaspi_2017_araa}, though still at least four orders-of-magnitude less energetic compared to supernovae and gamma-ray bursts. Nonetheless, FRBs display extreme brightness temperatures $T_{\rm B} \gg 10^{12}$\,K (the typically assumed threshold between incoherent and coherent emitters), and therefore must originate from a coherent emission process \citep{snelders_2023_natas}. This makes them detectable across cosmic distances despite their relatively modest energetics compared to other extreme astrophysical phenomena.

The distribution of observed FRB spectral energies ($E_{\nu}$) follow a differential power law, where the rate above some spectral energy scales as $R~(>E_{\nu}) \propto E_{\nu}^{\gamma_{D}}$ for a differential index $\gamma_{D}$. For the overall population of apparently non-repeating sources \citep{james_2022_mnras, shin_2023_apj}, $\gamma_{D}\sim-1.5$. Highly active repeating FRBs provide the opportunity to measure the burst energy distribution of a single source. In the case of repeaters, the energy distribution has been shown to deviate from a simple power law \citep{li_2021_natur, kirsten_2024_natas}. 

The most energetic FRBs are also by far the most rare. Therefore, large on-sky time is essential to probe the extremes of the FRB population. By probing the maximum energetics of FRBs, we can constrain the emission mechanism \citep{lu_2019_mnras, cooper_2021_mnras}; the total cosmic population \citep{luo_2020_mnras, james_2022_mnras, shin_2023_apj}; and inform how best to detect FRBs with upcoming telescopes.

There are various conceivable ways to investigate the maximum achievable energies and luminosities of FRBs, including 1. discovery of exceptionally distant or energetic one-off bursts \citep{ryder_2023_sci}; 2. population modelling of the full observed sample of FRBs \citep{james_2022_mnras, shin_2023_apj}; and 3. high-cadence monitoring of hyperactive repeating sources \citep{kirsten_2024_natas, sheikh_2024_mnras} (those repeaters that are sometimes seen to produce hundreds of bursts per hour, if observed with a high-sensitivity radio telescope). These methods complement each other, given that they are all subject to different observational biases and challenges. Moreover, it remains unclear whether apparently one-off and repeating FRBs share the same progenitors and emission mechanisms. By `progenitor', we mean the type of astrophysical source powering the bursts, and by `emission mechanism' we mean the physical process that generates the bursts. Probing the maximum observed energies of repeaters and apparently one-off FRBs is thus also a way to compare their nature.

In this work we investigate the maximum burst energy and luminosity of \roneoneseven, using multiple 25-32\,m--class radio telescopes that together provide unprecedented observational coverage in terms of on-sky time. \roneoneseven was discovered using the CHIME/FRB system \citep{mckinven_2022_atel}, and was soon identified as a hyperactive source compared to most other known repeaters. It was localised to a host galaxy at $z=0.0771$ \citep{ravi_2023_apjl,hewitt_2024_mnras} and has been the target of many follow-up observations. The exceptionally high activity of \roneoneseven makes it an ideal source to map the burst energy distribution of a repeater. Here we specifically focus on the maximum achievable burst energy and luminosity.

\section{Observations}

We observed \roneoneseven using four European radio telescopes: the Westerbork RT-1 25-m in the Netherlands (Wb); the Onsala 25-m in Sweden (O8); the Stockert 25-m in Germany (St); and the \torun\ 32-m in Poland (Tr). These observations span $117$~days between MJD $59867$ and $59983$ (2022 October 15 until 2023 February 08) for a total of $2192$~hours, which reduces to $1491$~hours of unique on-source time when taking into account the overlap between different observing modes and telescopes. During our campaign we observed at P-band ($330\,\mathrm{MHz}$), L-band ($1.4\,\mathrm{GHz}$), and C-band ($4.7\,\mathrm{GHz}$) with the aim of observing simultaneously over a wide bandwidth, as much as possible. Table~\ref{tab:coverage} gives an overview of the different observing setups while Figure~\ref{fig:obs_overview} visually represents the observations that were taken. Our data recording and burst searching strategy match that of earlier work \citep{kirsten_2024_natas}. At Wb, Tr and O8 we recorded amplitude and phase data (raw voltages) and at St we recorded total intensity data. We searched for bursts using standard methods adapted to the specifics of each telescope.  

\begin{table*} 
\caption{\label{tab:coverage}Observational set-up.}
\resizebox{\textwidth}{!}{%
\begin{tabular}{c c c c c c S[table-format=3.1] S[table-format=3.1] S[table-format=2.2] S[table-format=3.2,input-decimal-markers={.,}]}
\hline
{Station$\mathrm{^{a}}$}  & {Band} & {Frequency} & {Bandwidth$\mathrm{^{b}}$} & {Bandwidth per} & {SEFD$\mathrm{^{c}}$} & {Detection$\mathrm{^{d}}$} & {Completeness$\mathrm{^{e}}$} & {Completeness$\mathrm{^{f}}$} & {Time observed$\mathrm{^{g}}$} \\
 &  & {[MHz]} & {[MHz]} & {subband [MHz]} & {[Jy]} & {threshold [Jy~ms]} & {threshold [Jy~ms]} & {threshold [$\mathrm{10^{30}\,erg\,Hz^{-1}}$]} & {[hr]} \\
\hline
Wb  & P                 & 300--356          &50     & 8  & 2100  & 46.5 & 172.5 & 23.36   & 607.34\\
Wb  & L$_{\rm Wb}$      & 1259--1387        &100    & 16 & 420   & 6.6  & 24.4  & 3.30   & 196.58\\
St  & L$_{\rm St}$      & 1332.5--1430.5    &90     & 98 & 385   & 6.4  & 23.6  & 3.20   & 933.53 \\
Tr  & L$_{\rm Tr-1}$    & 1290--1546        &200    & 32 & 350   & 3.9  & 14.4  & 1.95   & 3.23 \\
Tr  & L$_{\rm Tr-2}$    & 1350--1478        &100    & 16 & 350   & 5.5  & 20.3  & 2.75   & 268.10 \\
Tr  & C$_{\rm Tr-1}$    & 4550--4806        &200    & 32 & 220   & 2.4  & 7.9   & 1.07   & 40.39 \\
Tr  & C$_{\rm Tr-2}$    & 4600--4728        &100    & 32 & 220   & 3.4  & 11.3  & 1.53   & 113.87 \\
O8  & C$_{\rm O8}$      & 4798.5--5054.5    &200    & 32 & 480   & 5.3  & 17.4  & 2.36   & 28.36 \\
\hline
\multicolumn{9}{l}{Total time at $1.4\,\mathrm{GHz}$ (L-band) on source [hr]$\mathrm{^{h}}$} & \textrm{1159} \\
\hline
\multicolumn{9}{l}{Total telescope time/total time on source [hr]$\mathrm{^{h}}$} & \textrm{2192/1491} \\
\hline

\multicolumn{8}{l}{$\mathrm{^{a}}$ Wb: Westerbork RT1, St: Stockert, Tr: \torun, O8: Onsala $\mathrm{25-m}$} \\
\multicolumn{8}{l}{$\mathrm{^{b}}$ Effective bandwidth accounting for RFI and band edges.} \\
\multicolumn{8}{l}{$\mathrm{^{c}}$ From the \href{https://www.evlbi.org/sites/default/files/shared/EVNstatus.txt}{EVN status page} (with the exception of St).} \\
\multicolumn{8}{l}{$\mathrm{^{d}}$ Assuming a $7\sigma$ detection threshold and a typical FRB pulse width of $1~\mathrm{ms}$.} \\
\multicolumn{8}{l}{$\mathrm{^{e}}$ Assuming a $15\sigma$ detection threshold and a width of $3~\mathrm{ms}$.} \\
\multicolumn{8}{l}{$\mathrm{^{f}}$ Fluence completeness converted using equation \ref{eq:energy}.} \\
\multicolumn{8}{l}{$\mathrm{^{g}}$ Hours spend on source between MJD 59867 and 59983 (2022 October 15 and 2023 February 08).} \\
\multicolumn{8}{l}{$\mathrm{^{h}}$ Total time on source accounts for overlap between the participating telescopes.} \\
\end{tabular}
}
\end{table*}

\subsection{Westerbork, Onsala \& \torun}\label{sec:sub_evn_dishes}

The data reduction and burst search analysis at Westerbork, Onsala and \torun follows our custom pipeline, which has been previously described \citep{kirsten_2024_natas}. 
At these three telescopes, we captured and stored the raw voltages (waveform data) in \texttt{.VDIF} format \citep{whitney_2010_ivs}, with dual circular polarisations and 2-bit sampling. In order to search for bursts, we converted the \texttt{.VDIF} data to 8-bit total intensity (Stokes~I) \sigproc filterbank files using \texttt{digifil} \citep{vanstraten_2011_pasa}. In order to limit dispersive smearing within a channel we made filterbanks with different time and frequency resolutions for searches at different observing bands. For C-band observations, we made filterbanks with $4$\,$\upmu$s time bins and $250$\,kHz wide frequency channels. For L-band observations at \torun this was $8$\,$\upmu$s time bins and $62.5$\,kHz channels and for Westerbork $64$\,$\upmu$s time bins and $62.5$\,kHz channels; while for P-band we used $512$\,$\upmu$s time bins and $7.8125$\,kHz channels.

We used \href{https://sourceforge.net/projects/heimdall-astro/}{{\tt Heimdall}} to search for bursts in the filterbank files. In order to minimise the amount of false positive candidates we set a signal-to-noise (S/N) threshold of $7$ and limit the dispersion measure (DM) search to within $\pm\,50$ \dmunit of the reported value of $\mathrm{DM_{FRB}=220}$~\dmunit \citep{mckinven_2022_atel}. We mitigate radio frequency interference (RFI) by applying a static mask, which excises certain frequency channels known to contain RFI. The identified burst candidates are then classified using the machine learning classifier \texttt{FETCH} \citep{agarwal_2020_mnras}, where we make use of models A $\&$ H and set a detection threshold of $50\,\%$. The burst candidates that have a reported probability of at least $50\,\%$ in one of the two models are then all manually inspected. As a fail-safe we also manually inspect all burst candidates that have a reported DM within $5$ \dmunit of the expected DM. 

Using the radiometer equation we calculate both the detection and completeness thresholds for the observations \citep{cordes_2003_apj}. The detection threshold represents the minimal fluence that our telescopes are sensitive to, while the completeness threshold represents the minimal fluence where we expect to detect (almost) all bursts. For the detection threshold we assume a 7-$\sigma$ detection with a canonical adopted FRB width of $1~\textrm{ms}$. For the completeness threshold we assume a conservative 15-$\sigma$ threshold and take a burst width of $3~\textrm{ms}$ since we know that bursts originating from \roneoneseven have generically longer durations than one millisecond \citep{konijn_2024_mnras}. The various detection and completeness thresholds of the instruments, per observing band, are listed in Table~\ref{tab:coverage}. 

\subsection{Stockert}\label{sec:sub_stockert}

At Stockert we record 32-bit total intensity data using the Pulsar Fast Fourier Transform (PFFTS) backend \citep{barr_2013_mnras}. These data are stored in \texttt{PFFTS} format. Using the \texttt{filterbank} tool from the \sigproc package, we create \texttt{filterbank} files that consist of 32-bit floats. The time- and frequency resolution of the filterbanks are $218.45$\,$\upmu$s and $586$\,kHz, respectively. We then search for bursts using tools from the \presto package \citep{ransom_2011_ascl}. The \texttt{rfifind} tool is used to mitigate RFI, \texttt{prepsubband} then dedisperses the data using a DM of $\mathrm{DM_{FRB}=220}$~\dmunit, and finally we search for burst candidates using \texttt{single\_pulse\_search} with a S/N limit of $8$. Reported burst candidates are then classified using \texttt{FETCH} Model~A and a detection threshold of $50\,\%$. All automatically classified burst candidates are then manually inspected. The detection and completeness threshold are listed in Table~\ref{tab:coverage}.

\subsection{Observational pointing}\label{sec:pointing}

Westerbork and \torun pointed towards RA=23$^{\rm{h}}$09$^{\rm{m}}$05\fs6 Dec=+48$^{\circ}$42$^{\prime}$00\farcs0 (J2000), as published in the discovery report by \citet{mckinven_2022_atel}. Onsala pointed towards RA=23$^{\rm{h}}$09$^{\rm{m}}$05\fs49 Dec=+48$^{\circ}$42$^{\prime}$25\farcs6 (J2000), the initial localisation as reported by DSA-110 \citep{ravi_2022_atel_15693}. Stockert initially also pointed towards the CHIME localisation, but changed this to the DSA-110 localisation after 21 December 2022. Even though there was an erratum for the DSA-110 localisation \citep{ravi_2022_atel_15716}, we did not change our pointing. Regardless, the offset between the pointing directions of the telescopes and the best-known position of \roneoneseven is less than $\sim30^{\prime\prime}$ in all cases \citep{hewitt_2024_mnras}. This pointing offset is still well within the full-width at half maximum primary beams at different wavelengths, which range from $0.1{\degr}$ (C-band) to $2.3{\degr}$ (P-band). 

\section{Analysis and Results}

We detected 130 bursts from \roneoneseven. Of these, 114 unique bursts were detected at $1.4\,\mathrm{GHz}$, including 16 bursts that were detected by multiple telescopes simultaneously. Remarkably, the highest-energy bursts from \roneoneseven contributed to $22.0^{+15.6}_{-10.3}\,\%$ of the all-sky rate of FRBs at L-band ($\mathcal{F}>500\mathrm{~Jy~ms}$). We also detected $16$ unique bursts at $330\,\mathrm{MHz}$, while no bursts were detected at $4.7\,\mathrm{GHz}$. Although we sometimes had simultaneous coverage between all observing bands, see Figure~\ref{fig:obs_overview}, we did not detect any burst at multiple frequency bands. Each burst is labeled with an ID, $\rm{B}x$, numbered in order of their arrival time, and followed by the telescope code to indicate which instrument detected the burst (e.g., B15-Tr). A subset of bursts, preferentially those with high S/N, is shown in Figure~\ref{fig:family_plot_subset}; a complete overview of all dynamic spectra is provided in the Supplementary material.

\begin{figure*}
    \centering
    \includegraphics[width=\textwidth]{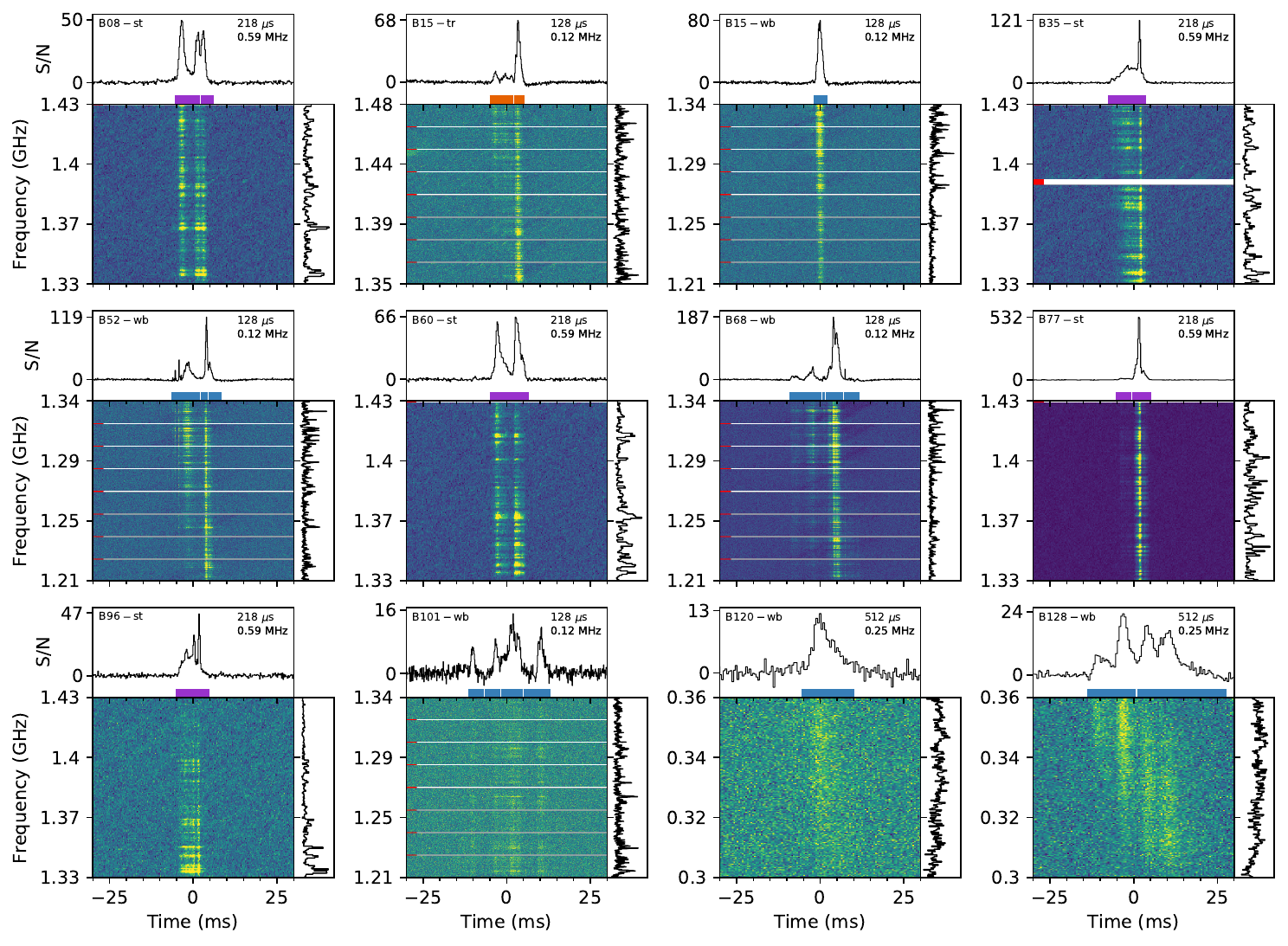}
    \caption{Dynamic spectra, time and temporal profiles for a subset of bursts. Each subfigure consists of three panels. Shown in the top panel is the burst-id, the time- and frequency resolution at which the data is plotted and the time profile of the burst. The colored bars represent the width for each component of a burst whereas the color of the bars correspond the instrument used to detect the burst. Purple corresponds to Stockert (St), orange to \torun (Tr) and blue to Westerbork (Wb). The side panel shows the temporal profile which is the sum over the time axis, but only under the colored bars. The white vertical lines are masked channels at the edges of the subbands or the presence of radio frequency interference (RFI) which are indicated by red ticks. The bursts have been corrected for dispersive effects where we used a value of $219.37$~\dmunit for bursts detected at $1.4~\mathrm{GHz}$ (L-band) and $219.73$~\dmunit for bursts detected at $0.3~\mathrm{GHz}$ (P-band). For Stockert this correction was applied incoherently (between frequency channels) and for \torun and Westerbork this correction was applied incoherently and coherently (within frequency channels). The dynamic spectra of all detected bursts are available as part of the Supplementary material.}
    \label{fig:family_plot_subset}
\end{figure*}

\subsection{Burst properties}

To correct for the dispersive delay we used a single DM for all bursts detected at L-band ($1.4\,\mathrm{GHz}$). The value we used is $219.375$\,\dmunit, which was determined by \citet{hewitt_2023_mnras} from analysis of the extremely bright and broad-band microshots in a burst that was co-detected by NRT (B2 in that paper) and Westerbork (B52-Wb here). While the DM of \roneoneseven is known to vary \citep{hewitt_2023_mnras}, we find that such variations have a negligible effect on the inferred fluences and hence energetics of the bursts, which is the main focus of this paper. Hence, we do not attempt to derive an optimal DM for each burst, but see \S\ref{sec:dm}.

The burst properties we use in our analyses were determined using the filterbank files created with \texttt{digifil} and the SPC-algorithm, previously referred to as Method~\rom{3}. RFI was mitigated by manually masking affected frequency channels for all bursts using the tools \texttt{psrzap} and \texttt{pazi} from the DSPSR software package. Additionally, we also zapped the edges of the subbands because of the drop in sensitivity at these frequencies (see, e.g., B35-st and B15-tr in Fig.~\ref{fig:family_plot_subset}).

To measure the time and frequency extent of the bursts we follow our earlier work \citep{kirsten_2024_natas}. We manually determined the start and stop times of each individual component of a burst. For these time ranges we calculate the 2D auto-correlation function in time and frequency. We then fit a 1D Gaussian to the spectra and time series and determine the width both in time and frequency. We define the burst frequency extent to be the full extent of the observing frequency if the full width at half maximum (FWHM) of the Gaussian fit is more than $75\,\%$ of the total bandwidth. The fluence of a burst is determined using the radiometer equation and calculating the flux density per time bin while summing over the on-time region for each component \citep{cordes_2003_apj}. An overview of burst properties can be found in Table~\ref{tab:burst-properties-ext}.

Additionally, we also measured the fluences using filterbank files created by SFXC (Method~\rom{1}) and only \texttt{digifil} (Method~\rom{2}) in order to compare and quantify the digitisation effects that are described in more detail in \S\ref{sec:digi_art}. The ratios of fluences between the three methods are shown in Figure~\ref{fig:fluence_ratio_band}. We find that we encounter saturation effects that underestimate the fluences by up to $40\,\%$ (SFXC) and overestimate them by $10\,\%$ (\texttt{digifil}) for bursts detected at L-band (left panel). For bursts detected at P-band (right panel) the saturation effect is less apparent even though we detect bright bursts ($>1000$\,Jy~ms). This lack of saturation is most likely due to the longer dispersive sweep of the burst and the relatively low sensitivity of the Westerbork P-band receiver.

When determining the time of arrival (ToA) of a burst one has to take into account potential data loss during the recording of the \texttt{.VDIF} format data. \texttt{digifil} currently has no functionality to account for potential data loss that occurred during an observation. Instead, when creating a filterbank file, \texttt{digifil} will stitch together gaps in the data. In order to correct for this issue, we determined the ToAs of our detected bursts using SFXC. SFXC has the functionality to accurately handle data loss by padding the missing data with zero values. We create coherent dedispersed filterbank files for bursts observed with Wb and Tr at a time resolution of $64 \ \upmu \textrm{s}$ for L-band bursts and $512 \ \upmu \textrm{s}$ for P-band bursts. For St, missing data is padded in real time. Both programs use the same assumed dispersive constant of $1/(2.41 \times 10^{-4})$\,MHz$^{2}$\,pc$^{-1}$\,cm$^{3}$\,s. We fit a Gaussian to every component in the time series for each burst. We define the ToA of a component as the centre of this fitted Gaussian and we set the ToA of a single burst as the centre of the Gaussian in case of a single component burst or the middle point between the left and right-most component in the case of a multi-component burst. For the case of SFXC, the timestamps are reported with respect to the geocenter of the Earth with the reference frequency being the middle of the top subband. For \texttt{digifil}, the timestamps are local arrival times and the reference frequency is the middle of the top frequency channel. For both cases, we convert the arrival times to barycentric arrival times in the TDB timescale with respect to infinite frequency for the assumed DM for L-band and P-band. An overview of all determined ToAs, per burst and per component, can be found in Table~\ref{tab:burst-properties-ext}.\\

\begin{table*}
\caption{Burst properties. The complete table and an additional table with properties per component for each burst can be found in the supplementary material.}
\label{tab:burst-properties-ext}
\resizebox{\textwidth}{!}{%
\begin{tabular}{c c S[table-format=5.12,group-digits=none] S[table-format=2.2] S[table-format=4.2(5),separate-uncertainty=true] c S[table-format=2.2] S[table-format=4.2(5),separate-uncertainty=true] S[table-format=4.2(5),separate-uncertainty=true] c c}
\hline
{Burst ID$^{\dagger}$} & Station & {TOA$^\mathrm{a}$} & {Peak S/N$^\mathrm{b}$} & {Fluence$^\mathrm{c}$} & {Number} & {Width$^\mathrm{d}$} & {Spectral density$^\mathrm{e}$} & {Spectral luminosity$^\mathrm{f}$} & BW$^\mathrm{g}$ & Central Frequency \\
 &  & {[MJD]} &  & {[Jy ms]} & {of components} & [ms] & {[$\mathrm{10^{29}\,erg\,Hz^{-1}}$]} & {[$\mathrm{10^{32}\,erg\,s^{-1}\,Hz^{-1}}$]} & [MHz] & [MHz] \\ \midrule
B01 & St & 59867.55026197426    & 14.90          & 42.16 \pm 8.43 & 2 & 9.61 & 57.11 \pm 11.42 & 5.94 \pm 1.19 & 128 & 1381 \\
B02 & St & 59868.72071901775    & 8.91          & 24.26 \pm 4.85 & 1 & 7.86 & 32.86 \pm 6.57 & 4.18 \pm 0.84 & 128 & 1381 \\
B03 & St & 59868.80699833712    & 4.69          & 29.73 \pm 5.95 & 1 & 13.54 & 40.27 \pm 8.05 & 2.97 \pm 0.59 & 128 & 1381 \\
B04 & St & 59868.95206450199    & 4.41          & 11.8 \pm 2.36 & 1 & 4.59 & 15.98 \pm 3.2 & 3.48 \pm 0.70 & 128 & 1381 \\
B05 & Tr & 59868.95206954337   & 16.67         & 40.75 \pm 8.15 & 1 & 3.97 & 55.19 \pm 11.04 & 13.91 \pm 2.78 & 128 & 1414 \\
B06 & Tr & 59868.98521494057    & 5.89          & 20.31 \pm 4.06 & 1 & 6.14 & 27.51 \pm 5.5 & 4.48 \pm 	0.90 & 128 & 1414 \\
{$\vdots$} & $\vdots$           & {$\vdots$}    & {$\vdots$} & {$\vdots$} & {$\vdots$} & {$\vdots$} & {$\vdots$} & {$\vdots$} & {$\vdots$} & {$\vdots$} \\
B127 & Wb & 59941.68738958076   & 6.99          & 274.25\pm 54.85 & 1 & 20.48 & 371.47 \pm 74.29 & 18.14 \pm 3.63 & 54 & 328 \\
B128 & Wb & 59941.73017555207  & 24.17         & 2629.4 \pm 525.88 & 4 & 32.77 & 3561.49 \pm 712.3 & 108.71 \pm 21.74 & 54 & 328 \\
B129 & St & 59949.88269849737   & 6.78          & 46.15 \pm 9.23 & 1 & 12.67 & 62.51 \pm 12.5 & 4.93 \pm 0.99 & 128  & 1381 \\
B131 & St & 59978.53235768391   & 39.33         & 375.07 \pm 75.01 & 1 & 13.33 & 508.03 \pm 101.61 & 38.13 \pm 7.63 & 128 & 1381 \\
B132 & Tr & 59982.71918144583   & 22.65         & 50.59 \pm 10.12 & 1 & 6.91 & 68.53 \pm 13.71 & 9.92 \pm 1.98 & 128 & 1414 \\ \bottomrule
\multicolumn{10}{l}{$\mathrm{^{\dagger}}$B84 was omitted due to it later being classified as radio frequency interference (RFI), B130 was removed because the presence of strong RFI made it impossible to measure its properties.} \\
\multicolumn{10}{l}{$\mathrm{^{a}}$Time of arrival referenced to the solar system barycenter at infinite frequency in TDB. We used a dispersion constant of  $1/(2.41 \times 10^{-4})$\,MHz$^{2}$\,pc$^{-1}$\,cm$^{3}$\,s.} \\
\multicolumn{10}{l}{For detections with a central frequency of $328$~MHz we used a DM of $219.735$ \dmunit and the others a DM of $219.37$ \dmunit.} \\
\multicolumn{10}{l}{$\mathrm{^{b}}$The peak S/N of the brightest component.} \\
\multicolumn{10}{l}{$\mathrm{^{c}}$The sum of the computed fluences for each component measured after applying the SPC algorithm (method~\rom{3}).} \\
\multicolumn{10}{l}{We assume a $20\%$ error for all bursts dominated by the uncertainty of the SEFD. } \\
\multicolumn{10}{l}{$\mathrm{^{d}}$The manually determined time span between the start of the first and end of the last component.} \\
\multicolumn{10}{l}{$\mathrm{^{e}}$Computed using Equation \ref{eq:energy}, $D_L=362.4$~Mpc and $z=0.0771$.} \\
\multicolumn{10}{l}{$\mathrm{^{f}}$Spectral density divided by the width.} \\
\multicolumn{10}{l}{$\mathrm{^{g}}$The bandwidth of the burst used for the computation of the fluence.} \\
\end{tabular}%
}
\end{table*}

\subsection{Detection and sky rate} \label{sec:disc_skyrate}

We observed \roneoneseven for  $1158.57$ and $607.34$ unique hours at L- and P-band, respectively, and detected $13$ and $7$ bursts above a fluence $\mathcal{F}>500$~Jy~ms. This implies a detection rate of $0.27^{+0.19}_{-0.13}$~burst/day (L-band) and $0.28^{+0.29}_{-0.17}$~burst/day (P-band) for high-fluence bursts, where the errors are the 95\% Poisson uncertainty on the rates. Though the burst rates are consistent between the two bands, we caution that these multi-frequency observations were not strictly simultaneous and both reflect an average rate over some range of time (Figure~\ref{fig:obs_overview}). 

Based on the ASKAP Fly Eye's survey \citep{shannon_2018_natur} and modelled number counts \citep{lu_2019_apj} assuming a burst energy distribution with slope $\gamma_{\rm{C}}=-1.5$, an all-sky rate at L-band has been determined above a fluence $\mathcal{F}>100\mathrm{~Jy~ms}$, $\mathrm{R_{sky}}(\mathcal{F}>100~\mathrm{Jy~ms})=5\times10^3~\mathrm{sky^{-1}~yr^{-1}}$. We detected $48$ L-band bursts above $100$~Jy~ms, which corresponds to a rate of $7.3^{+2.4}_{-1.9}\,\%$ $\mathrm{R_{sky}}(\mathcal{F}>100\mathrm{~Jy~ms})$ and agrees with the all-sky rate for the ATA sample of $5.8^{+3.4}_{-2.4}\,\%$ \citep{sheikh_2024_mnras} (quoted errors are the $95\%$ Poisson uncertainty on the rates). For our brightest detections, $13$ bursts above $500$~Jy~ms and $4$ bursts above $1000$~Jy~ms, we find an all-sky rate of $22.0^{+15.6}_{-10.3}\,\%$ $\mathrm{R_{sky}}(\mathcal{F}>500\mathrm{~Jy~ms})$ and $19.1^{+29.8}_{-13.9}\,\%$ $\mathrm{R_{sky}}(\mathcal{F}>1000\mathrm{~Jy~ms})$, respectively. A similar calculation has been done for the all-sky rate contribution for \rsixseven with a fluence larger than $500$~Jy~ms and was estimated to be $2.6^{+5.0}_{-2.0}\,\%$ $\mathrm{R_{sky}}(\mathcal{F}>500\mathrm{~Jy~ms})$ \citep{kirsten_2024_natas}. This illustrates that the contribution to the all-sky rate of \roneoneseven was almost an order of magnitude higher compared to \rsixseven for highly energetic bursts. Additionally, it underlines that a single hyperactive FRB repeater can strongly contribute to the all-sky rate for a relatively short time when the source is active. The upcoming, wide-field BURSTT \citep{lin_2022_pasp} telescope (FoV\,$\sim 10^{4}~\rm{deg}^{2}$) should be an excellent system for identifying such sources.

\subsection{Cumulative burst rates}

In addition to the new data we present here, our analysis also makes use of three additional observational campaigns towards \roneoneseven, which were performed during the same time range: $8.67$\,h of observations with the Five-hundred-meter Aperture Spherical Telescope \citep[FAST;][]{zhang_2023_apj}; $61$\,h of observations with the \nancay Radio Telescope \citep[NRT;][]{konijn_2024_mnras}; and $541$\,h of observations with the Allen Telescope Array \citep[ATA;][]{sheikh_2024_mnras}. 

The cumulative burst distribution of FRBs is sometimes fit by a single power law $R~(>E_{\nu}) \propto E_{\nu}^{\gamma_{C}}$, as has been done, e.g., for \rone\citep{gourdji_2019_apjl} and  \meightone\citep{nimmo_2023_mnras}. Here $R$ is the rate of bursts, $E_{\nu}$ the spectral energy density of the bursts, and $\gamma_{C}$ is the slope of the cumulative distribution. To be able to compare distributions between telescopes, we express the energetics of the bursts as spectral energy density (erg\,Hz$^{-1}$), $E_{\nu} = E / \nu$. Where $\nu$ is the observed bandwidth of the burst. We convert the measured fluences to spectral energy via \citep{macquart_2018_mnras},
\begin{equation}
E_{\nu} = \frac{\mathcal{F} \cdot 4 \pi D_{L}^2}{(1+z)^{2+\alpha}}
\label{eq:energy}
\end{equation}    
where $\mathcal{F}$ is the fluence of the burst, $4 \pi D_{L}^2$ is the luminosity distance factor assuming isotropic emission, and $(1+z)^{2+\alpha}$ is the redshift correction, where $\alpha$ is the spectral index ($\mathcal{F}_{\nu} \propto \nu^{\alpha}$). We set $\alpha = 0$ to be consistent with respect to the fluence calculation. The fluence is calculated based on the band-averaged time series where we thus assume $\alpha = 0$. The luminosity distance ($D_{L}$) of \roneoneseven is $362.4$\,Mpc, with corresponding redshift $z=0.0771$ \citep{ravi_2022_mnras}.

To fit a power law to the cumulative distribution, we exclude bursts that were detected below our completeness threshold for St, Wb, and Tr; see Table~\ref{tab:coverage} for an overview of completeness thresholds per telescope and observing band. 
The sensitivity of NRT and FAST enables the detection of bursts of much lower spectral energy density ($E_{\mathrm{min}}\sim~10^{28}-10^{29}$\,erg\,Hz$^{-1}$) where the cumulative distribution has been shown to deviate from a single power law \citep{li_2021_natur}. In this work, we focus on a turnover towards the higher energies. Therefore, we use the Python package \texttt{powerlaw} \citep{Alstott_2014_PLoSO} to determine the minimum energy ($E_{\rm min}$) above which the distribution is best described by a power law. For FAST we find $E_{\rm min}^{\rm{FAST}} = 5.6\times10^{29}$\,erg\,Hz$^{-1}$ and for NRT we find $E_{\rm min}^{\rm{nc}} = 3.4\times10^{29}$\,(erg\,Hz$^{-1}$). 

An initial guess of the power law index is estimated using a maximum likelihood method \citep{crawford_1970_apj,james_2019_mnras}. Next, we use \texttt{scipy.optimize.curvefit} to fit a power law, where we assume a fiducial uncertainty of $20\,\%$ on the energy of the burst. This uncertainty stems mainly from the uncertainty on the system equivalent flux density (SEFD) of each telescope. The error from \texttt{curvefit} is the first error we quote. In addition to the $1\sigma$ error quoted by \texttt{curvefit} we perform a bootstrapping method to estimate the variance of the fit. Bootstrapping is done by refitting the data with only a subset of the data points. We refitted the data 1000 times using $90\,\%$ of the data points without replacement. The error determined via bootstrapping is noted as the second error quoted on the derived values. 

\begin{figure*}
    \includegraphics[width=\textwidth]{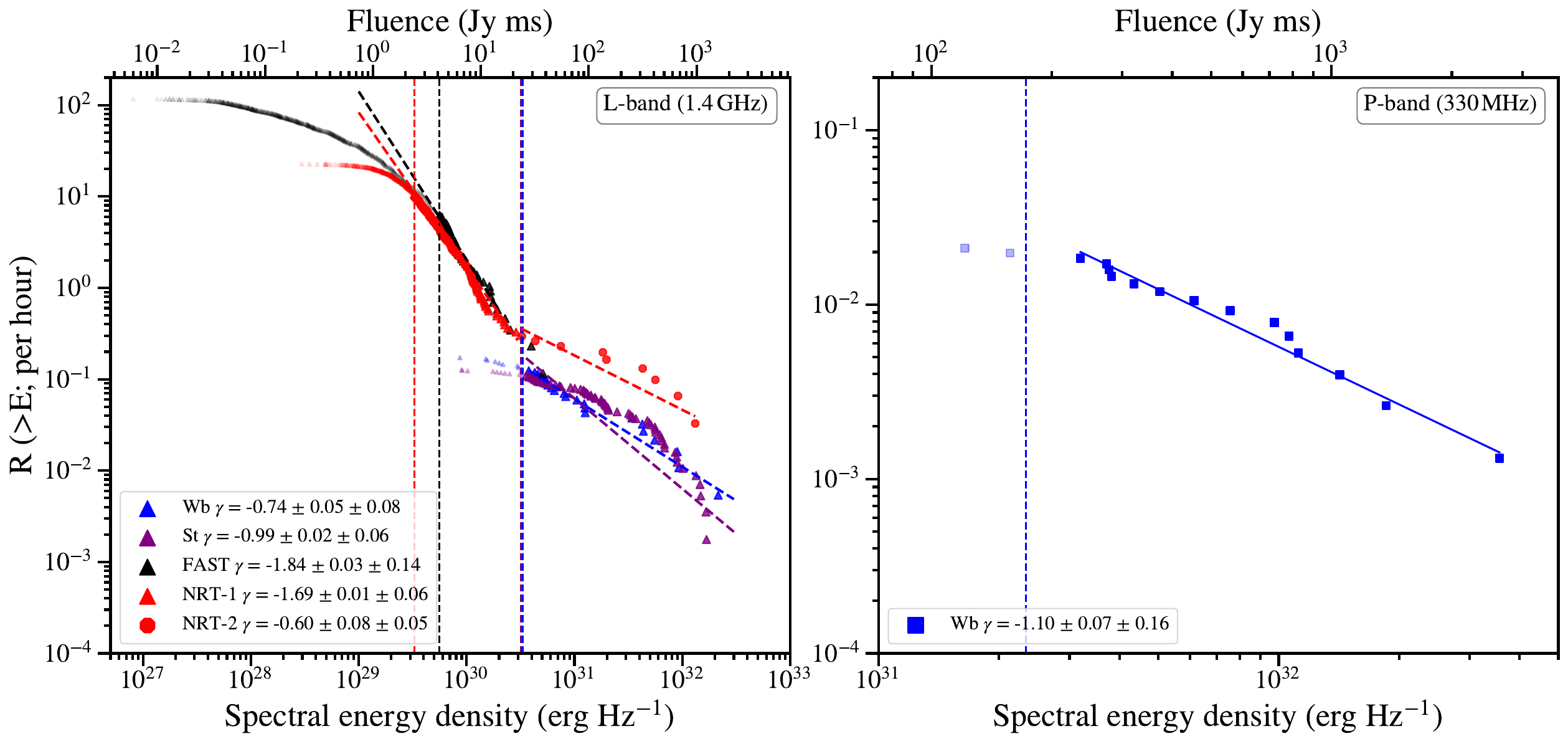}
    \caption{Cumulative burst energy distribution of spectral energy densities.  In the left panel we show detections by Westerbork (Wb) and Stockert (St), FAST \citep{zhang_2023_apj} and NRT \citep{konijn_2024_mnras} at $1.4~\mathrm{GHz}$ (L-band). In order compare between the different observational campaigns we only show bursts that were observed between MJD 59869 and 59910. Comparing the different rates reveals a break in the distribution towards higher energies ($\sim3\times10^{30}$\,erg\,Hz$^{-1}$). The purple and blue vertical line correspond to the completeness threshold as indicated in Table \ref{tab:coverage}. The red and black vertical lines denote the point where the distribution can be best described by a single power law as calculated by the Python package \texttt{powerlaw}. Transparent data points which are on the left side of the vertical lines were excluded in the fit. When fitting we set a $20\,\%$ error on the energies and quote two errors. The first error is the $1\sigma$ statistical uncertainty on the fit and the second error is the $1\sigma$ error after the bootstrapping method. In the right panel we show detections observed at $0.3~\mathrm{GHz}$ (P-band).}  
    \label{fig:burst_rate_pl_band}
\end{figure*}

In Figure~\ref{fig:burst_rate_pl_band} we show the cumulative burst energy distribution, denoted with slope $\gamma_{C}$, for detections from Westerbork, Stockert, FAST and NRT at L-band ($1.4\,\mathrm{GHz}$) --- between MJD~$59869$~(2022 October 17) and MJD~$59910$~(2022 November 27). Constraining the time period allows us to directly compare the determined slopes for the different observational campaigns, while avoiding the potential pitfall of a burst energy distribution that evolves with time. For the high-energy bursts ($E_{\nu}>3\times10^{30}$\,erg\,Hz$^{-1}$) detected by Stockert we find a power-law index of $\gamma_{C}^{\mathrm{st}} = -0.99\pm0.02\pm0.06$ and for Westerbork we find $\gamma_{C}^{\mathrm{Wb}} = -0.74\pm0.05\pm0.08$. For the low-energy bursts ($E_{\nu}<3\times10^{30}$\,erg\,Hz$^{-1}$) from FAST we find a significantly steeper slope of $\gamma_{C}^{\mathrm{FAST}} = -1.84\pm0.03\pm0.13$. Figure~\ref{fig:burst_rate_pl_band} also shows the cumulative distribution for the bursts detected at P-band ($330\,\mathrm{MHz}$) by Westerbork, where we find a best-fit power law with index $\gamma_{C}^{\mathrm{Wb}} = -1.10\pm0.07\pm0.15$. 

\begin{figure*}
    \centering
    \includegraphics[width=\textwidth]{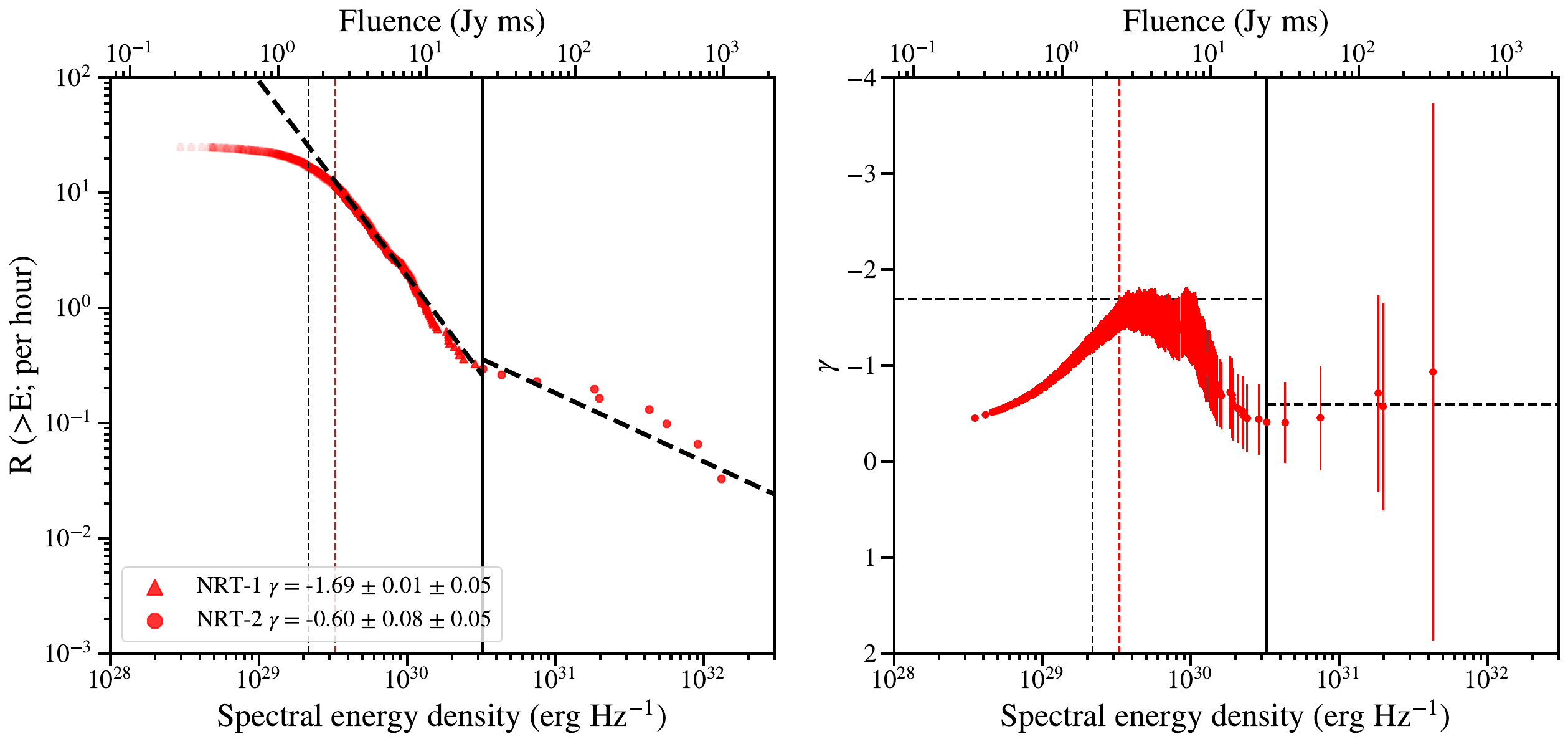}
    \caption{Cumulative burst energy distribution for bursts detected by NRT. Left: The cumulative energy distribution for NRT, also shown in Figure~\ref{fig:burst_rate_pl_band}, fitted with two power laws. The vertical dotted black line is the estimated completeness threshold and the vertical dotted red line indicates the turnover point as estimated using the \texttt{powerlaw} package. The solid black line corresponds to the determined breakpoint of the distribution at $E_{\rm{break}} = 3.2~\times~10^{30}$\,erg\,Hz$^{-1}$. Right: the power law index ($\gamma_{\rm{C}}$) as a function of the spectral energy density calculated by a maximum-likelihood estimation. The horizontal dotted black lines indicate the slopes of the power law used in the left panel.}
    \label{fig:nc_two_slopes}
\end{figure*}

We find that the burst energy distribution of NRT is not well described by a single power law. This is apparent by eye, and by calculating the power-law slope using the maximum likelihood method as a function of the spectral energy density, as shown in the right panel of Figure~\ref{fig:nc_two_slopes}. We therefore fit a broken power law to the distribution and determine the breakpoint to be $E_{\rm{break}}\sim3.2\times10^{30}$\,erg\,Hz$^{-1}$. To be consistent in our methodology, we fit two separate power laws to the NRT data. For the first power law, for bursts that satisfy $E_{\rm min}^{\rm{NRT}} < E_{\nu}^{\rm{NRT}} < E_{\rm{break}}$, we find $\gamma_{C}^{\mathrm{NRT-1}} = -1.69\pm0.01\pm0.05$. We fit a second power law for bursts that satisfy $E_{\rm{break}} < E_{\nu}^{\rm{nc}}$ and find $\gamma_{C}^{\mathrm{NRT-2}} = -0.60\pm0.08\pm0.05$, as shown in the left panel of Figure~\ref{fig:nc_two_slopes}. Interestingly, the integrated total energy of the bursts in the low- and high-energy regimes of the energy distribution are roughly equal.

\subsection{Turnover at the characteristic energy} \label{sec:result_turnover}

\begin{figure*}
    \centering
    \includegraphics[width=\textwidth]{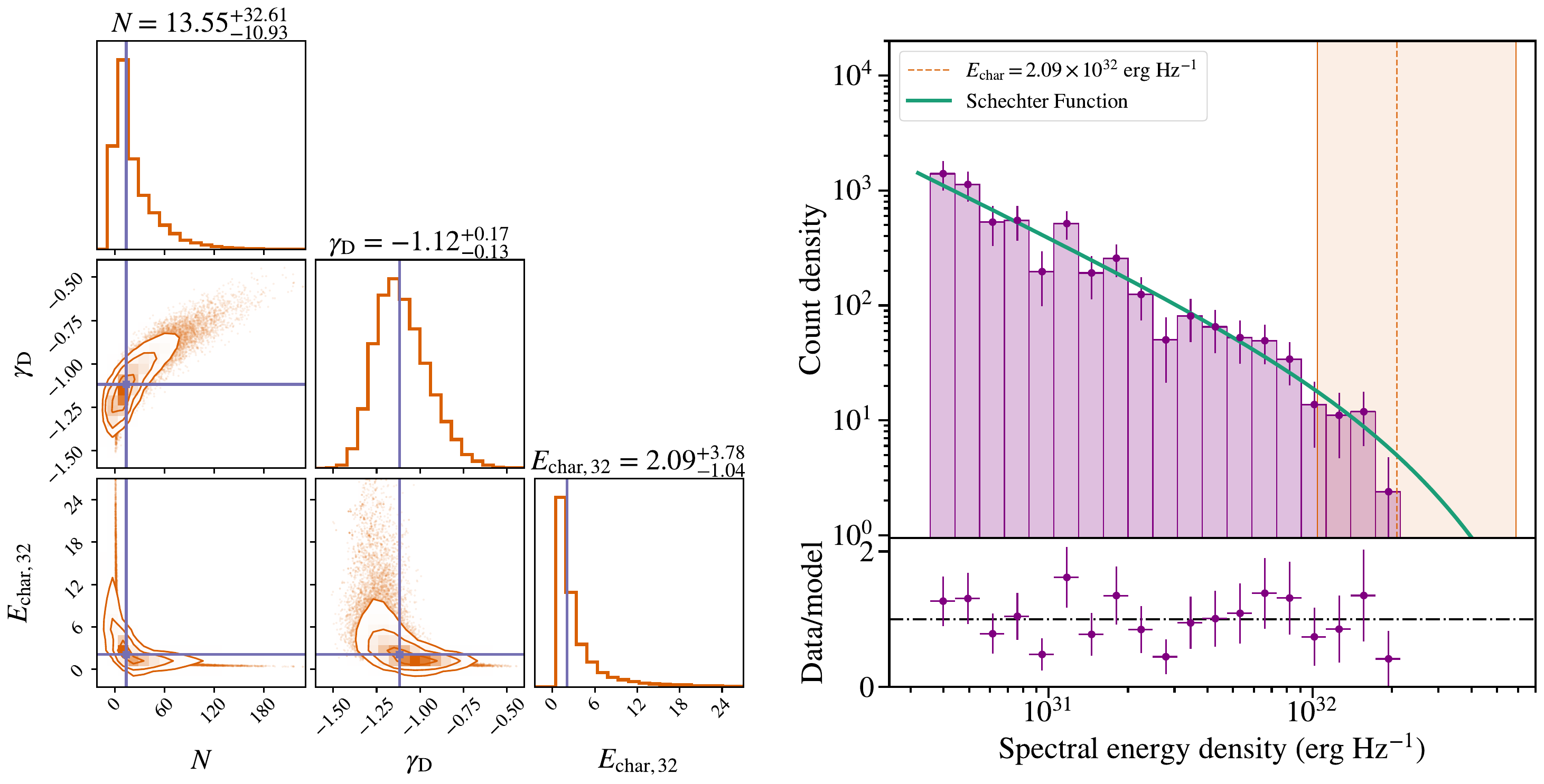}
    \caption{\textbf{Spectral energy densities modelled by a Schechter function.} \\Left: corner plot of the results from the MCMC analysis, thinned by a factor $30$ for visual purposes. The solid lines denote the median values of the posterior distributions and the errors are the $16\,\%$ and $84\,\%$ quantile. Right: The differential distribution in the top panel includes bursts above $3.3~\times~10^{30}$\,erg\,Hz$^{-1}$ (or $24.4$~Jy~ms) detected by Westerbork, Stockert, \torun, NRT, ATA and FAST between MJD 59869 and 59910. The green line is an over-plotted Schechter function based on the median values of the posterior distribution of the MCMC analysis. The vertical orange dashed line is the best fit value for the characteristic energy with the coloured region corresponding to error region of the $16\,\%$ and $84\,\%$ quantile. The bottom panel shows the ratio between the data and the Schechter model.
    }
    \label{fig:MCMC_schechter_fig1}
\end{figure*}

As we expect a physical limit to FRB energies we investigate a possible turnover in the energy distribution. We combined detections from multiple observational studies constrained in time between MJD 59868 and MJD 59910. These studies were the NRT sample \citep{konijn_2024_mnras}, detections from ATA \citep{sheikh_2024_mnras} and FAST \citep{zhang_2023_apj}, as well as the detections described in this work. We filtered out $5$ duplicate bursts that were co-detected by the ATA and Stockert and only used the brightest detection in each case. We only consider bursts with energies larger than the completeness threshold of the least sensitive telescope (Westerbork at $24.4$\,Jy~ms), see Table~\ref{tab:coverage}. We assume a Schechter function \citep{schechter_1976} as,
\begin{equation}
    P(E_{\nu}) = N \left( \frac{E_{\nu}}{E_{\textrm{char}}} \right)^{\gamma_{D}} \textrm{exp} \left[- \frac{E_{\nu}}{E_{\textrm{char}}}  \right]
\end{equation}
where $E_{\nu}$ is the spectral energy density (erg\,Hz$^{-1}$), $\gamma_{D}$ is the slope on the differential distribution, $E_{\textrm{char}}$ is the cut-off energy, and $N$ is a normalisation factor. We apply a Markov Chain Monte Carlo (MCMC) fitting technique to test if the combined burst sample is well described by a Schechter function. We bin the sample of bursts into 20 independent bins, and infer the model parameters using a Bayesian model comprised of a Poisson likelihood function appropriate for binned count data, and flat priors on the parameters. We set the flat prior on $E_\mathrm{char}$ to lie between 0 and 1000 in units of $10^{32}$erg\,Hz$^{-1}$. We sampled the model using MCMC as implemented in \texttt{emcee} \citep{Foreman_emcee} with $100$ walkers and $10000$ steps after $500$ burn-in steps. A corner plot of the posterior probability densities generated with the \texttt{corner} package \citep{Foreman_corner_py} is shown in the left panel of Figure \ref{fig:MCMC_schechter_fig1}. Tests using narrower priors on $E_\mathrm{char}$ between $0.5$ and $50$ in units of $10^{32}$erg\,Hz$^{-1}$ produced similar results, indicating robustness against changes in prior density.

We find a characteristic maximum energy of $E^{\textrm{char}}_{\nu}=2.09^{+3.78}_{-1.04}~\times~10^{32}$\,(erg\,Hz$^{-1}$) with a differential power law index of $\gamma_{D}=-1.12^{+0.17}_{-0.13}$ that we directly compare to previous modelling of one-off FRBs in Table \ref{tab:sch_fit_compare}. In Figure~\ref{fig:MCMC_schechter_fig1} we show the differential distribution of burst energies over-plotted with a Schechter function using the median values of the posterior distributions from the MCMC run. We also indicate the determined characteristic maximum energy and uncertainty range.

\begin{table}
\caption{\label{tab:schec_comp} Best Schechter fit results from different studies.}
\resizebox{\columnwidth}{!}{%
\begin{tabular}{l c c c}
\hline
{}  & {$\textrm{log}_{10}~E^{\textrm{char}}$ [erg]$^\mathrm{a}$}  & {$\gamma_{D}^\mathrm{b}$} &  {$\mathcal{F}_{\textrm{min}}$[Jy ms]$^\mathrm{e}$} \\ \midrule
{This work}                            & $41.32^{+0.45}_{-0.30}$   & $-1.12^{+0.17}_{-0.13}$   & 24.4 \\ 
{\citet{ryder_2023_sci}$^{,\mathrm{c}}$}  & $41.7^{+0.2}_{-0.2}$      &  {...}                    & {...} \\
{\citet{shin_2023_apj}}                & $41.38^{+0.51}_{-0.50}$   & $-1.30^{+0.7}_{-0.4}$     & 5.0 \\
{\citet{james_2022_mnras}}             & $41.26^{+0.27}_{-0.22}$   & $-1.95^{+0.18}_{-0.15}$   & 0.5/4.4/21.9$^\mathrm{f}$ \\
{\citet{luo_2020_mnras}$^{,\mathrm{d}}$}  & $42.08^{+0.30}_{-0.06}$   & $-1.79^{+0.31}_{-0.35}$   & 0.55 \\ \bottomrule
\multicolumn{4}{l}{$\mathrm{^{a}}$Energy assuming a canonical adopted bandwidth of $1$~GHz.} \\
\multicolumn{4}{l}{$\mathrm{^{b}}$Best fitted slope parameters on a differential distribution.} \\
\multicolumn{4}{l}{$\mathrm{^{c}}$Based on the sample of \citet{james_2022_mnras}.} \\
\multicolumn{4}{l}{$\mathrm{^{d}}$Luminosity ($L_{*}$) converted using the average width from Table 2 of $\bar{w}=4.14~\mathrm{ms}$.} \\
\multicolumn{4}{l}{$\mathrm{^{e}}$Fluence limits on the different observational campaigns.} \\
\multicolumn{4}{l}{$\mathrm{^{f}}$From \citet{james_2022_mnras_z_dm}.} \\
\end{tabular}%
}
\label{tab:sch_fit_compare}
\end{table}

\section{Discussion}\label{sec:discussion}

\subsection{Activity of hyperactive repeaters} \label{sec:disc_hyper}

We find that the burst energy distribution of \roneoneseven can not be described by a single power law. Rather, it is well described by a broken power law ($E^{\rm{break}}_{\nu} \sim 3.2~\times~10^{30}$\,erg\,Hz$^{-1}$) and an exponential cut-off ($E^{\rm{char}}_{\nu} = 2.09^{+3.78}_{-1.04}~\times~10^{32}$\,erg\,Hz$^{-1}$). The burst energy distribution is steeper ($\gamma_{C}^{\mathrm{NRT-1}} = -1.69\pm0.01\pm0.05$) below the break, and flatter ($\gamma_{C}^{\mathrm{NRT-2}} = -0.60\pm0.08\pm0.05$) above it. Previous work also showed that another hyperactive repeater, \rsixseven, has a similar break in its energy distribution at ($E^{\rm{break}}_{\nu} \sim 8~\times~10^{30}$\,erg\,Hz$^{-1}$) as well as a similar flattening: from $\gamma_{C}^{\mathrm{FAST}} = -1.95\pm0.001\pm0.06$ at lower energies to $\gamma_{C}^{\mathrm{O8+St}} = -0.48\pm0.11\pm0.03$ at the highest burst energies \citep{kirsten_2024_natas}. The occurrence of a break requires explanation; it could indicate two populations of bursts produced by distinct physical mechanisms \citep{kirsten_2024_natas}. Approximately a similar amount of total energy is released by the source above and below the break. The similarity in the break point energy between \roneoneseven and \rsixseven suggests that this is a common feature that may be observed in the future for other repeaters. 

For \rsixseven, mapping the energy distribution required comparing high-energy bursts from high-cadence observations with 25-32\,m-class telescopes \citep{kirsten_2024_natas} with lower-energy burst detections from FAST \citep{xu_2022_natur}. For \roneoneseven we follow the same approach, comparing our detections to the lower-energy bursts observed by FAST \citep{zhang_2023_apj}. Additionally, we are able to detect the break in the power law burst energy distribution using NRT data alone \citep{konijn_2024_mnras} because NRT has good sensitivity and ample exposure time during the most intense period of activity (Figures~\ref{fig:burst_rate_pl_band} and \ref{fig:nc_two_slopes}; in comparison, the smaller dishes have a lower apparent rate because they missed the highest activity window around MJD~59879). The NRT distribution initially follows the slope of FAST, but then breaks and flattens towards higher energies following the slope of St and Wb. The observation of a break in the distribution by a single instrument is unprecedented; it underlines and confirms the flattening of the energy distribution towards higher energies. 

The ATA observed \roneoneseven for $541$\,h and fit a power law to the cumulative distribution of burst energies \citep{sheikh_2024_mnras}. They find a slope of $\gamma^{\rm ATA}_{C}=-1.08^{+0.25}_{-0.25}$. The energy range we are able to probe in this work overlaps with the ATA and allows for direct comparison. We find that the slopes for different instruments beyond the break point in the distribution, see Figure~\ref{fig:burst_rate_pl_band}, are consistent within uncertainties between all telescopes. 

The telescopes used in this work (St, Wb, Tr, and O8) all have observing bandwidths ranging between $56-256$\,MHz, which is considerably narrower compared with the ATA ($672$\,MHz) and NRT ($512$\,MHz). Comparing the results of the cumulative power-law indices between all telescopes in the high-energy range shows comparable results, which indicates that the flattening in the distribution towards higher energies can not simply be accounted for by the limited bandwidth of observation.

We observed $14$ bursts at P-band above our completion threshold, see Figure~\ref{fig:burst_rate_pl_band}. Based on our sample we are unable to shed light on whether the cumulative energy distribution breaks at the same energy compared to L-band and if the differential distribution turns over at a similar characteristic energy. To further investigate the presence of a break-point at low radio frequencies, requires a higher-sensitivity observational campaign capable of probing bursts in the lower-energy regime ($E_\nu<10^{30}$\,erg\,Hz$^{-1}$). 

\subsection{Characteristic maximum energy} \label{sec:disc_char}

Determining the characteristic maximum spectral energies of FRBs is crucial to constrain their emission mechanisms and nature. For \roneoneseven we find $E^{\textrm{char}}_{\nu}=2.09^{+3.78}_{-1.04}~\times 10^{32}$\,erg\,Hz$^{-1}$ or, equivalently, a total energy $\textrm{log}_{10} (E^\textrm{char}) = 41.32^{+0.45}_{-0.30}$\,erg, assuming a 1-GHz emission bandwidth (Figure~\ref{fig:MCMC_schechter_fig1}). Our findings agree with the inferred fluence limit $\mathcal{F} \lesssim 10^4$\,Jy~ms \citep{sheikh_2024_mnras} (or, equivalently, $\textrm{log}_{10} (E^\textrm{char}) = 42.15$\,erg) for \roneoneseven to be consistent with the all-sky fluence distribution extrapolated from ASKAP Fly's Eye survey \citep{shannon_2018_natur}.

Previously, constraints on $E^\textrm{char}$ have been derived from population studies of apparently one-off FRBs. In these cases, the energy distribution is modelled as a Schechter function. Two of these studies combined data from various surveys including UTMOST, HTRU and CRAFT \citep{luo_2020_mnras}, with another study only using the FRBs in the first CHIME/FRB catalog \citep{shin_2023_apj}. The detection of an FRB at $z = 1.016$ with an implied burst energy of $6.4 \pm 0.7~\times~10^{41}$\,erg also enabled a strong constraint on the characteristic maximum energy \citep{james_2022_mnras}. From these various studies, $E^\textrm{char}$ ranges between $\textrm{log}_{10} (E^\textrm{char}) = 41.26 - 42.08$\,erg (Table~\ref{tab:schec_comp}). Our determined value of $E^\textrm{char}$ for \roneoneseven is consistent with the different population studies of apparently one-off FRBs. 

However, in this comparison we have assumed a fiducial $1$\,GHz emission bandwidth for both repeaters and (apparent) non-repeaters alike. While the broad-band spectra of FRBs are still poorly constrained in general, previous studies with CHIME/FRB have found a systematic difference in the emission bandwidths of repeaters versus (apparent) non-repeaters \cite{pleunis_2021_apj}. Given this, our $\textrm{log}_{10} (E^\textrm{char})$ value could be a factor of $2-3\times$ lower if \roneoneseven bursts are restricted to $300-500$\,MHz emission bandwidths. While previous studies have found restricted emission bandwidth for \roneoneseven \citep{sheikh_2024_mnras,konijn_2024_mnras} in the $\sim 1-2$\,GHz range, its instantaneous emission bandwidth across the entire radio frequency range is unknown. Thus, for simplicity, and given the lack of a better model, we scale our total energies to a common 1-GHz emission bandwidth for both repeaters and (apparent) non-repeaters.

For the slope of the differential distribution, we find $\gamma_{D} = -1.12^{+0.17}_{-0.13}$, which agrees with previous studies though the uncertainties on the measured slopes are large (Table~\ref{tab:schec_comp}). In our fitting of a Schechter function, we have only considered bursts above the completeness threshold of the least sensitive telescope (Wb) and with $E^{\rm char}_{\nu} > 3.3\times 10^{30}$\,erg\,Hz$^{-1}$, to avoid bursts that follow a steeper power-law index below this value (Fig.~\ref{fig:burst_rate_pl_band}). We note that previous population studies may be including low-energy bursts that follow a different energy distribution, and which would skew the power-law slope of a Schechter fit to be apparently steeper. 

During our burst search, we set the detection threshold at S/N $> 7$. However, in our previous work \citep{kirsten_2024_natas}, we adopted a conservative approach and assumed a S/N of $15$ as our completeness threshold. In this work, we test the dependence of the characteristic maximum energy on the adopted completeness threshold by varying the assumed S/N. Varying the threshold involves modelling of the Schechter function with different total number of bursts, see Supplementary Table \ref{tab:MCMC-runs}. The completeness thresholds are referenced with respect to the completeness of the least sensitive telescope (Wb). For a threshold of $10\sigma$, the reduced chi-square implies a poor fit, which can be explained by our sample of detections from the 25$-$32-m telescopes being incomplete at that threshold. Adopting higher detection thresholds of $20-$ and $30\sigma$ results in $\chi^2<1$, which indicates overfitting of the data. Placing the completeness threshold exactly at the breakpoint ($E_{\rm{break}}$) of the distribution gave results consistent with those for a $15\sigma$ threshold, suggesting that this threshold appropriately characterises our observations. We fit the Schechter function to the burst energy distribution, which was grouped into 20 independent bins. Running the same analysis with 15 and 25 bins produced consistent results, demonstrating that the fits do not strongly depend on the chosen binning.

\begin{table*}
\caption{Schechter function parameter results. Results and best fit parameters from the MCMC analysis on fitting a Schechter function to the differential burst distribution by varying the fluence/energy threshold for bursts detected between MJD 59869 and 59910. The top row denotes the the completeness threshold and parameter values which were used in Figure \ref{fig:MCMC_schechter_fig1}.}
\label{tab:MCMC-runs}
\resizebox{\textwidth}{!}{%
\begin{tabular}{c S[table-format=2.2] S[table-format=1.2] c c c c c c c c c c S[table-format=1.2]}
\hline
{$\sigma$~Threshold} & {fluence Threshold} & {spectral energy Threshold} & {Total Bursts} & {$N_{\textrm{Wb}}$} & {$N_{\textrm{St}}$} & {$N_{\textrm{Tr}}$} & {$N_{\textrm{ATA}}$} & {$N_{\textrm{Nc}}$} & {$N_{\textrm{Fast}}$} & {N} & {$\gamma_{D}$} & {$E^{\textrm{char}}$} & {$\chi^{2}_{\nu}$}\\ 
{} & {{[Jy ms]}} & {[$\mathrm{10^{30}\,erg\,Hz^{-1}}$]} & {} & {} & {} & {} & {} & {} & {} & {} & {} & {[$\mathrm{10^{32}\,erg\,Hz^{-1}}$]} & {}\\ \midrule
15$^{\mathrm{b}}$  & 24.40             & 3.30      & 125   & 22    & 62    & 2     & 8     & 29    & 2    & $13.55^{+32.61}_{-10.93}$  & $-1.12^{+0.17}_{-0.13}$ & $2.09^{+3.78}_{-1.04}$ & 0.78 \\ \midrule
10                 & 16.10             & 2.18      & 146   & 25    & 68    & 4     & 13    & 32    & 4    & $9.68^{+24.87}_{-8.30}$ & $-1.14^{+0.14}_{-0.10}$ & $2.54^{+6.87}_{-1.33}$ & 1.13 \\
14.55$^\mathrm{a}$ & 23.81             & 3.22      & 130   & 23    & 64    & 2     & 9     & 30    & 2    & $12.44^{+28.52}_{-10.14}$  & $-1.13^{+0.16}_{-0.12}$ & $2.16^{+4.22}_{-1.06}$ & 0.85 \\
20                 & 32.45             & 4.40      & 112   & 20    & 55    & 2     & 7     & 27    & 1    & $20.32^{+40.40}_{-16.08}$  & $-1.05^{+0.20}_{-0.15}$ & $1.68^{+2.70}_{-0.76}$ & 0.66 \\
30                 & 48.80             & 6.61      & 93    & 12    & 47    & 1     & 7     & 26    & 0    & $40.06^{+66.02}_{-31.28}$ & $-0.90^{+0.28}_{-0.23}$ & $1.19^{+1.68}_{-0.49}$ & 0.50 \\ \bottomrule
\multicolumn{13}{l}{$\mathrm{^{a}}$Energy threshold corresponding the the break-point of the power law determined for NRT, see Figure~\ref{fig:nc_two_slopes}.} \\
\multicolumn{13}{l}{$\mathrm{^{b}}$The completion threshold corresponding to the least sensitive telescope (Wb). The result of this fit are shown in Figure~\ref{fig:MCMC_schechter_fig1}.}\\
\end{tabular}%
}
\end{table*}

To test whether a Schechter function is preferred over a simpler power-law function, we performed an identical MCMC analysis for a power-law model. This allowed us to compute the Bayesian information criterion (BIC), which is defined as:
\begin{equation} \label{eq:bic_equation}
    \rm{BIC} = k\ln{(n)} - 2\ln{(\hat{L})}
\end{equation}
Where $k$ is the number of parameters, $n$ the number of data points and $\hat{L}$ is the sample with the highest likelihood from an MCMC chain. The number of data points is in both models taken as the number of bins ($n=20$); the number of parameters is $k=2$ for the power-law function and $k=3$ for the Schechter function. Since we make use of flat priors, we take the sample with the highest probability as the maximum likelihood sample, which is $\ln(\hat{L})=-41.64$ for the power-law function and $\ln(\hat{L})=-33.53$ for the Schechter function. This results in BIC values of $89.26$ for the power-law function and $76.05$ for the Schechter function. A BIC difference of $13.21$ indicates that the fitting of the Schechter function is a more appropriate description of the data. This is because the burst energies abruptly cut off around $2\times10^{32}$\,erg\,Hz$^{-1}$, even though our observations could have detected bursts above this threshold.

\subsection{Comparison to energy distributions of pulsars} \label{sec:sub_pulsars}

Pulsars produce microsecond-to-millisecond duration coherent radio pulses, which are observationally similar to FRBs; they are hence a useful point of reference, even if the energy scales and progenitors of FRBs are different in nature. Some pulsars emit giant pulses (GPs) which show a burst distribution that could be described with a break and flattening toward higher energies. For PSR~B1937$+$21 a break occurs in the cumulative distribution for pulses with a pulse energy larger than $7~\textrm{Jy}~\upmu \mathrm{s}$ \citep{mckee_2019_mnras}. These results were not found in previous studies on this pulsar \citep{cognard_1996_apjl, soglasnov_2004_apj}, which could be due to the much shorter exposure times. the cumulative distribution of PSR~B0540$-$69 does show a flattening in the distribution for higher energetic bursts \citep{geyer_2021_mnras}.
Another well-studied emitter of GPs is PSR~B0531$+$21 (the Crab Pulsar). The term `supergiant pulse' was coined for pulses that show inconsistency with the probability implied by a simple power law \citep{cordes_2004_apj}. These findings of a flattening were confirmed after $100$ hours of observing with the Green Bank telescope \citep{mickaliger_2012_apj}. A more recent study did not find any hint of flattening in the distribution of supergiant pulses ($\mathcal{F}>130$~Jy~ms) after observing the Crab pulsar for $260$ hours \citep{bera_2019_mnras}.
Although we see similarities between FRBs and pulsars in their cumulative energy distributions, their energies differ by at least four orders of magnitude. A shared rotationally powered emission mechanism between pulsars and FRBs is ruled out; it is more likely that FRBs are magnetically powered \citep{Lyutikov_2017_ApJL}. 
To our knowledge, systematic flattening of pulsar energy distributions at high energies has not been studied in detail for a population of sources.

\subsection{Constraining the FRB emission mechanism} \label{sec:disc_emission}

Extremely bright FRBs are valuable for constraining the progenitors and emission mechanisms of FRBs. To examine these constraints, we consider the most energetic bursts observed at L-band and P-band: B68-Wb and B128-Wb, respectively. These bursts have measured energies of $2.8 \times 10^{40}$\,erg and $2.0 \times 10^{40}$\,erg, respectively, determined by multiplying the observed spectral energy density with the bandwidth of each burst. To be more conservative, this calculation differs from previous calculations where we use a canonical adopted bandwidth of $1$\,GHz to convert to burst energies. As such, these values can be considered as lower limits on the isotropic-equivalent burst energy.

For typical radio efficiencies assumed in `far-away' FRB models \citep{metzger_2019_mnras} of $\epsilon_{\rm radio} \approx 10^{-5}$, these bursts require extreme event energies $E_{\rm flare} > 10^{45}$\,erg, consistent with magnetar giant flares. Magnetospheric `close-in' FRB models often invoke crustal dislocations as the mechanism by which particle acceleration is triggered. Given an event energy, a required magnetic field can be derived assuming fiducial crustal oscillation parameters \citep{wadiasingh_2020_apj}. If $\epsilon_{\rm radio} \approx 10^{-5}$, an FRB with $E_{\rm FRB} = 2 \times 10^{40} \, {\rm erg}$ implies a local field strength $B \approx 2 \times 10^{16} \, {\rm G}$, suggesting that B68-Wb/B128-Wb require very large magnetic fields, plausibly stemming from multipolar field components or extreme crustal dislocations. We note that this magnetic field requirement corresponds to the specific mechanism presented in \cite{wadiasingh_2020_apj}. 

Coherent curvature radiation (CCR) by bunches has been discussed as a radiation mechanism for generating magnetospheric FRBs \citep{kumar_2017_mnras,wang_magnetospheric_2020}. There is a maximum peak luminosity of CCR based on a upper limit to the electric field $E_{\parallel}$, above which Schwinger pairs rapidly screen the field (\citealt{lu_maximum_2019}; see also \citealt{kumar_bosnjak_2020}). The requirement that the momenta of radiating particles remains well-aligned also implies a maximal bunch luminosity \citep{cooper_2021_mnras}.

\begin{figure}
    \centering
    \includegraphics[width=\columnwidth]{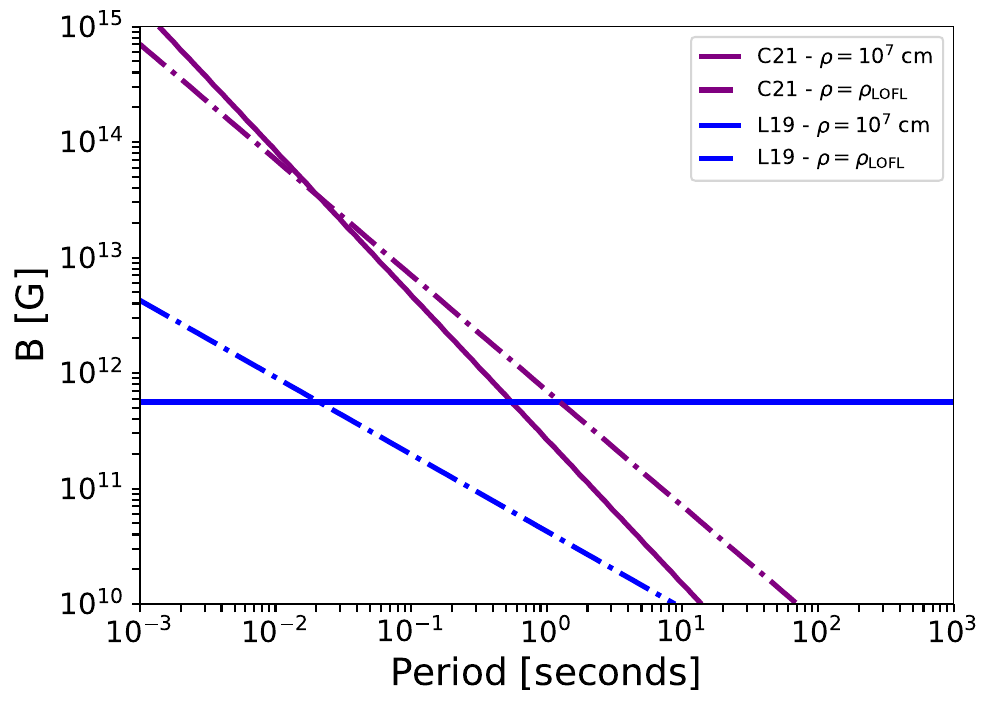}
    \caption{Magnetic field constraints for B77-St. Minimum local magnetic field $B$ as a function of neutron period for B77-St assuming coherent curvature radiation \citep{lu_maximum_2019,cooper_2021_mnras}. Solid lines refer to fixed field line curvature radius of $\rho_{\rm c} = 10^{7} \, {\rm cm}$, dot-dashed lines refer to emission along the last open field line (LOFL) at the polar cap co-latitude.}
    \label{fig:magnetospheric_limits}
\end{figure}

To characterise the limit of these constraints on the source of \roneoneseven, we consider the burst in our sample with the highest peak luminosity ($S_\nu$): B77-St. We calculate the luminosity as \citep{macquart_2018_mnras},
\begin{equation} \label{eq:luminosity}
    L_{\nu} = \frac{S_{\nu} \cdot 4 \pi D_{L}^2}{(1+z)^{1+\alpha}}
\end{equation} 
Where $S_{\nu}$ is the peak flux density, $4\pi D_{L}^2$ the distance luminosity factor and $(1+z)^{1+\alpha}$ the redshift correction. resulting in a peak spectral luminosity of $L_{\nu, \rm B77-st} = 1.38 \times 10^{35} \; {\rm erg \, s^{-1} \, Hz^{-1}}$, measured at a time resolution of $\Delta \rm{t}=218.45~\upmu s$. In Figure~\ref{fig:magnetospheric_limits}, we show the minimum local magnetic field of the emission region based on the aforementioned maximum luminosities. These limits set conservative constraints on the surface magnetic field (e.g., corresponding to emission directly from the surface). Therefore, if B77-St is powered by CCR along open field lines, it must stem from a region with $B \gtrsim 10^{12}\,$G.

\subsection{Source energy reservoir constraints} \label{sec:disc_resevoir}

Assuming a perfect dipolar magnetic field, the total external magnetospheric energy reservoir of a neutron star can be crudely estimated as:
\begin{equation}
    U_{\rm B} \approx \frac{R_{\rm NS}^3 B_{\rm s}^{2}}{6} \approx 1.7 \times 10^{47} \, {\rm erg} \; \bigg(\frac{R_{\rm NS}}{10^{6} \, {\rm cm}}\bigg)^3 \, \bigg(\frac{B_{\rm s}}{10^{15} \, {\rm G}}\bigg)^2
\end{equation}
Here, $B_{\rm s}$ is the surface magnetic field and $R_{\rm NS}$ is the neutron star radius. To examine the total energetics we integrate over the best fit differential burst energy distribution function (Fig.~\ref{fig:MCMC_schechter_fig1}). The total radio energy of bursts observed across the $1769$ unique observing hours of all observing campaigns at L-band included in the dataset (assuming minimal overlap) above $E_{\rm radio, min} = 3.3 \times 10^{30} \; {\rm erg \, Hz^{-1}}$ is: $1.3 \times 10^{42}$\,erg or $7.5 \times 10^{38} {\rm erg \, hr^{-1}}$ (assuming a conservative emission bandwidth $\Delta \nu \approx 100 \, {\rm MHz}$, given our limited observing bandwidth). If we naively assume that the luminosity beaming enhancement and radio bursts missed due to beaming cancel out, as well as a radio efficiency factor of $\epsilon_{\rm radio} = 10^{-5}$, we find the external magnetic energy will be depleted on a timescale:

\begin{equation}
\tau \approx 2150 \: {\rm hr} \; \bigg(\frac{B_{\rm s}}{10^{15} {\rm G}}\bigg)^2 \, \bigg(\frac{\epsilon_{\rm radio}}{10^{-5}}\bigg)
\end{equation}

From Figure~\ref{fig:burst_rate_pl_band}, we can see that a roughly equal amount of integrated energy is released $E_{\rm radio, min} < 3.3 \times 10^{30}$. This estimate is in good agreement with the estimated integrated power of lower energy bursts observed by FAST \citep{zhang_2023_apj} if a similar $\epsilon_{\rm radio}$ is assumed, despite caveats to the approximation including the non-uniformity of the true burst rate and the unknown beam1ing characteristics. This result implies that if a magnetar powers \roneoneseven, the radio efficiency may be $\epsilon_{\rm radio} > 10^{-5}$, the external magnetic field may be $B > 10^{15}$ G, the external magnetic field is continuously replenished (e.g., via core field expulsion), or strong multipolar field components power most of the emitted energy.
These findings are in agreement with the conclusions of \citet{zhang_2025_arxiv}, who find that the cumulative energy of over $11,000$ bursts detected from  \ronefourseven\ nearly deplete the energy stored in the magnetic field of a typical magnetar.

\subsection{Energetics and multi-wavelength counterparts}

Multi-wavelength counterparts to FRBs are expected within both magnetospheric \citep{cooper_2021_mnras, yang_2021_apj} and particularly maser shock models \citep{metzger_2019_mnras}. Simultaneous X-ray counterparts, as observed for the FRB-like burst from SGR~1935+2154 \citep{ridnaia_2021_natas,tavani_2021_natas,li_2021_natas} and prompt optical afterglows from FRB-associated flares \citep{lyutikov_2016_apjl,kilpatrick_2021_apjl,cooper_2022_mnras} are the most likely detectable counterparts to extragalactic FRBs. 

Constraints on high-energy counterparts have been placed \cite{pearlman_2023_natas} for bursts from the closest repeating extragalactic FRB source: FRB~20200120E at just 3.6 Mpc \citep{bhardwaj_2021_apjl}. The most energetic FRB from this source ($E_{\rm r, B4} = 2.8 \times 10^{33} \, {\rm erg}$) was detected simultaneous to NICER observations, broadly ruling out any simultaneous giant or intermediate magnetar X-ray flares \citep{pearlman_2023_natas}. The most energetic bursts presented in this work are roughly 7 orders of magnitude larger in total radio energy, at a distance approximately 100 times greater meaning, for similar radio efficiencies, multi-wavelength fluxes coincident with the brightest bursts presented here should be a factor $\sim 10^{3}$ times higher. This bolsters the case for sensitive X-ray observations of prolific repeaters known to produce very bright bursts, notwithstanding the challenging large required time on source.

\begin{figure}
    \centering
    \includegraphics[width=\columnwidth]{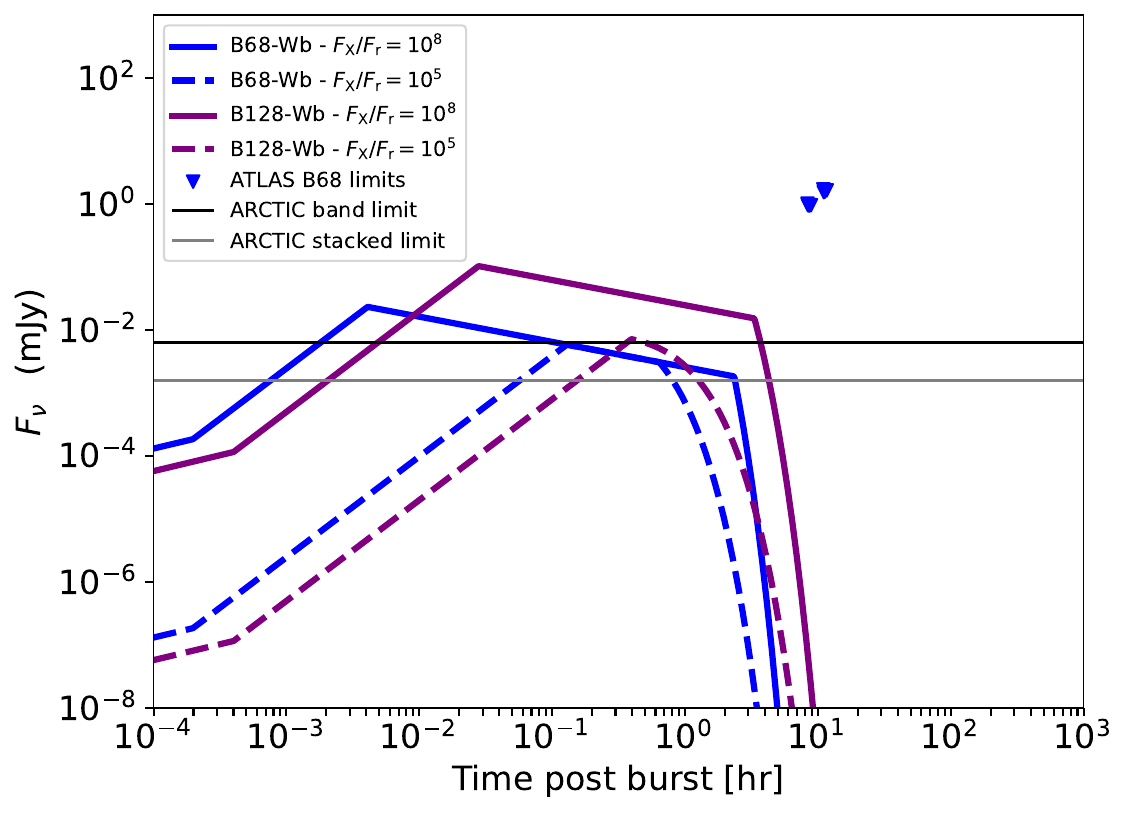}
    \caption{Predicted optical afterglows for maser shock model. The predicted optical afterglow from B68-Wb and B128-Wb in the synchrotron maser model framework \citep{cooper_2022_mnras}. We also show 5$\sigma$ limits obtained by the ATLAS telescope network from 8 hours after B68-Wb, and include typical limiting magnitudes from the ARCTIC telescope which could have been achieved if observations were scheduled rapidly after B68-Wb and B128-Wb \citep{kilpatrick_2021_apjl}.}
    \label{fig:optical_counterpart}
\end{figure}

In Figure~\ref{fig:optical_counterpart} we show the predicted approximate optical (680\,nm) afterglow lightcurves of B68-Wb and B128-Wb following previous maser shock afterglow formulations \citep{margalit_2020_apjl,cooper_2022_mnras}. In lieu of an X-ray detection, we normalize the lightcurves assuming an unobserved, quasi-simultaneous X-ray counterpart a factor $L_{\rm X}/L_{\rm r}$ more luminous than the FRB. We also show 5$\sigma$ upper limits obtained by the ATLAS telescope network \citep{tonry_2018_pasp} the night following B68-Wb in the `orange' $560-820$\,nm band, and include representative limiting magnitudes of the ARCTIC telescope \citep{huehnerhoff_2016_spie}, which has previously been utilized to constrain optical FRB afterglows \citep{kilpatrick_2021_apjl}.

\subsection{Prospects for detecting more high-energy FRBs} 

Following a method employed in earlier work \citep{kirsten_2024_natas}, we can place a limit on the maximum observable redshift based on our brightest detections at both $1.4$~GHz and $330$~MHz. Our brightest detection at $1.4$~GHz is burst B68-Wb with a reported fluence of $1587$~Jy~ms (or $E_\nu=2.2^{+0.4}_{-0.4}\times10^{32}$\,erg\,Hz$^{-1}$). FAST is the most sensitive telescope currently operating at this wavelength range, with a completeness threshold of $54$~mJy~ms \citep{xu_2022_natur}. B68-Wb would therefore still been detected by FAST if \roneoneseven would have been at a redshift of $z=9.9^{+1.1}_{-1.1}$ and the burst was emitted at $\sim14$~GHz. 

At $330$~MHz our brightest detection is burst B128-Wb with a measured fluence of $2629$~Jy~ms (or $E_\nu=3.6^{+0.7}_{-0.7}\times10^{32}$\,erg\,Hz$^{-1}$). With the reported detection limit of CHIME/FRB of $1$~Jy~ms \citep{mckinven_2022_atel}, burst B128-Wb would still have been observable if the source was placed at redshift $z=2.8^{+0.3}_{-0.3}$ with an implied emitting frequency of $\sim1.2$~GHz. Although the majority of CHIME sources do not have measured redshifts, their highest measured DM source, FRB~20180906B\xspace, is consistent with originating in the aforementioned redshift range with an estimated redshift of $z=2.95$ \citep{chawla_2022_apj}.
CHIME is not detecting a large population of these high-redshift sources \citep{shin_2023_apj}. This could be related to the large scattering for high-redshift FRBs ($z > 1$) at the observing frequency of CHIME \citep{ocker_2022_apj}.

These high-energy bursts, emitted at high redshift, could potentially be observed by future telescopes with good field-of-view (FoV) and increased sensitivity, such as CHORD \citep{vanderlinde_2019_clrp}. Upcoming radio telescopes with extremely large FoV but lower sensitivity, such as BURSTT \citep[FoV\,$\sim 10^{4}~\rm{deg}^{2}$;][]{lin_2022_pasp}, will accumulate a large number of observing hours on repeating sources like \roneoneseven and probe the high-energy burst distribution. Furthermore, these events could also be detected in the far side lobes of CHIME/FRB \citep{lin_2024_apj}. 

\section{Conclusions}

We conducted a high-cadence observational campaign towards \roneoneseven spanning $117$~days for a total of $2192$~hours between 2022 October and 2023 February. We detected a total of $130$ high-energy FRBs ($\mathcal{F}>10$\,~Jy~ms). Of these, $114$ unique bursts were detected at $1.4$\,GHz (L-band) and we detected $16$ bursts at $330$~MHz (P-band). Our main conclusions can be summarized as: 

\begin{itemize}
    \item We observe a break ($E^{\rm{break}}_{\nu} \sim 3.2~\times~10^{30}$\,erg\,Hz$^{-1}$) in the cumulative burst energy distribution with a flattening of the power-law slope at higher energies. Interestingly, a similar break has been observed in the energy distribution of \rsixseven by \cite{kirsten_2024_natas}.
    \item 
    We find a characteristic maximum energy at $E^{\textrm{char}}_{\nu}=2.09^{+3.78}_{-1.04}~\times 10^{32}$\,erg\,Hz$^{-1}$ or, equivalently, a total energy $\textrm{log}_{10} (E^\textrm{char}) = 41.32^{+0.45}_{-0.30}$\,erg. This value is consistent with measurements of the characteristic energy for different population studies of apparently one-off FRBs which range between $\textrm{log}_{10} (E^\textrm{char}) = 41.26 - 42.08$\,erg (Table~\ref{tab:schec_comp}). This could suggest a common physical mechanism for the emission of repeating and non-repeating FRB sources. Furthermore, we find that the integrated energy release above and below the break are approximately equal.
    \item When it was active, \roneoneseven contributed significantly to the all-sky FRB rate: $7.3^{+2.4}_{-1.9}\,\%$ $\mathrm{R_{sky}}(\mathcal{F}>100\mathrm{~Jy~ms})$ and $22.0^{+15.6}_{-10.3}\,\%$ $\mathrm{R_{sky}}(\mathcal{F}>500\mathrm{~Jy~ms})$. This fraction of the all-sky FRB rate was almost an order of magnitude higher than that of the other well-characterized hyperactive repeating source, \rsixseven.
    \item Based on the total observed radio energy at $1.4$~GHz (L-band), we are able to estimate that the total magnetic energy of a typical magnetar would be depleted on a time scale of $\tau \sim 2150$\,h. This time scale implies that the radio efficiency may be larger than $\epsilon_{\rm radio} \gg 10^{-5}$, the magnetic field may be $B \gg 10^{15}$\,G, the external magnetic field is continuously replenished (e.g., via core field expulsion), or strong multipolar field components power most of the emitted energy.
    \item We find that \roneoneseven can produce bursts that are energetic enough for optical telescopes to detect a fiducial afterglow of shock-based FRB models with prompt ($<$hour) follow-up (Fig.~\ref{fig:optical_counterpart}) and we show that the most energetic burst would have been detectable even if the source was at redshift $z=9.9^{+1.1}_{-1.1}$, assuming that high-energy bursts are emitted at higher radio frequencies.   
\end{itemize}

High-cadence monitoring totalling 100+ hours, not just telescope sensitivity, is essential for detecting the most bright and rare FRBs and revealing the high-energy tails of the energy distributions of (repeating) FRBs. Future observing campaigns similar to the one described in this work promise to provide additional insights on FRB sources and emission mechanisms.

\section*{Acknowledgements}

We thank the directors and staff of the participating telescopes for allowing us to observe with  their facilities. 
We thank Willem van Straten for modifying the DSPSR software package to fit our needs and for helping us with the SPC-algorithm. 
We thank Phil Uttley for discussions regarding MCMC analysis and fitting of the energy distribution.
This work makes use of data from the Westerbork Synthesis Radio Telescope owned by ASTRON. ASTRON, the Netherlands Institute for Radio Astronomy, is an institute of the Dutch Scientific Research Council NWO (Nederlandse Oranisatie voor Wetenschappelijk Onderzoek). We thank the Westerbork operators Richard Blaauw, Jurjen Sluman and Henk Mulder for scheduling and supporting observations.
This work is based in part on observations carried out using the 32-m radio telescope operated by the Institute of Astronomy of the Nicolaus Copernicus University in \torun (Poland) and supported by a Polish Ministry of Science and Higher Education SpUB grant. 
We express our gratitude to the operators and observers of the Astropeiler Stockert telescope: Thomas Buchsteiner, Bert Engelskirchen, Elke Fischer, Hans-Peter Löge, Thomas Nitsche and Kevin Schmitz.
This work is supported by the NWO XS grant: WesterFlash (OCENW.XS22.1.053; PI: Kirsten). 
The AstroFlash research group at McGill University, University of Amsterdam, ASTRON, and JIVE is supported by: a Canada Excellence Research Chair in Transient Astrophysics (CERC-2022-00009); the European Research Council (ERC) under the European Union’s Horizon 2020 research and innovation programme (`EuroFlash'; Grant agreement No. 101098079); and an NWO-Vici grant (`AstroFlash'; VI.C.192.045).
A.~J.~C. acknowledges support from the Oxford Hintze Centre for Astrophysical Surveys which is funded through generous support from the Hintze Family Charitable Foundation. 
D.~H. acknowledges support from NWO's Women In Science Excel (WISE) programme.
F.~K. acknowledges support from Onsala Space Observatory for the  provisioning of its facilities/observational support. The Onsala Space Observatory national research infrastructure is funded through Swedish Research Council grant No 2017-00648.
K.~N. is an MIT Kavli Fellow.
Z.~P. is supported by an NWO Veni fellowship (VI.Veni.222.295).

\newpage
\section*{Data Availability}

The data supporting the plots and full analysis in this article and other findings of this study are available under \url{https://doi.org/10.5281/zenodo.11261763}. The scripts and Jupyter notebooks used to analyse the data, generate the plots and tables with the burst properties are available at \url{https://github.com/astroflash-frb/frb20220912a-ouldboukattine-2025}. 


\newpage
\bibliographystyle{mnras}



\newpage
\appendix

\section{Observational overview}

\begin{figure*}
    \centering
    \includegraphics[width=\textwidth]{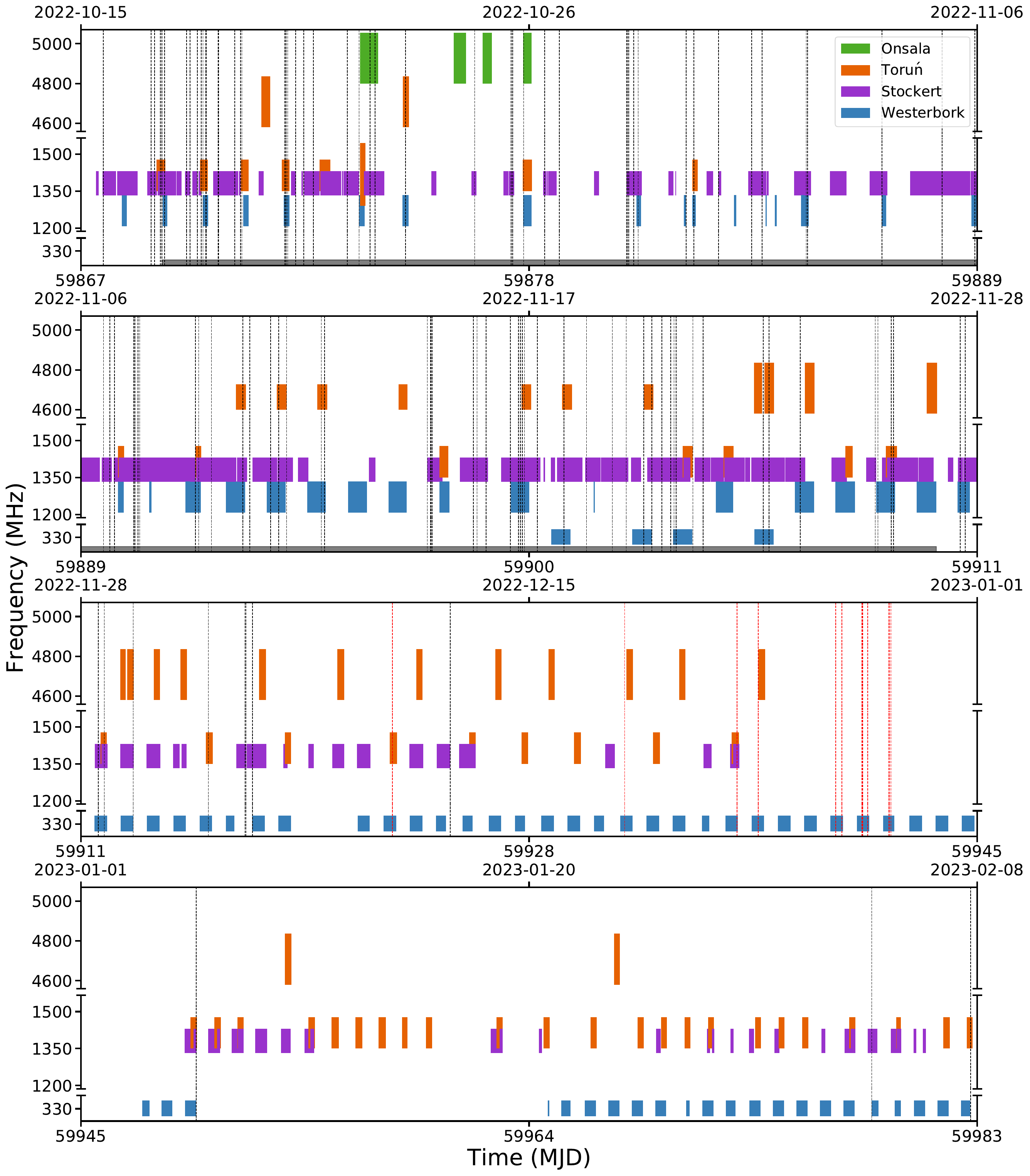}
    \caption{Overview of the \roneoneseven observing campaign. Each coloured block represents an observation with a telescope at a certain frequency. The top two panels show $22$ days and the bottom two panels show $36$ days with the associated MJD and and calendar dates on the x-axis. The broken y-axis show the frequency range. The vertical dotted lines denote detections of bursts where black illustrates detections at L-band ($1.4~\mathrm{GHz}$) and red detections at P-band ($0.3~\mathrm{GHz}$). The black bar in the top panels indicate the overlapping observing window with FAST and NRT.}
    \label{fig:obs_overview}
\end{figure*}

Observational overview of the \roneoneseven observing campaign is provided in Figure~\ref{fig:obs_overview}.

\section{Burst properties}

\subsection{Dispersion measures} \label{sec:dm}

\begin{figure*}
    \centering
    \includegraphics[width=\textwidth]{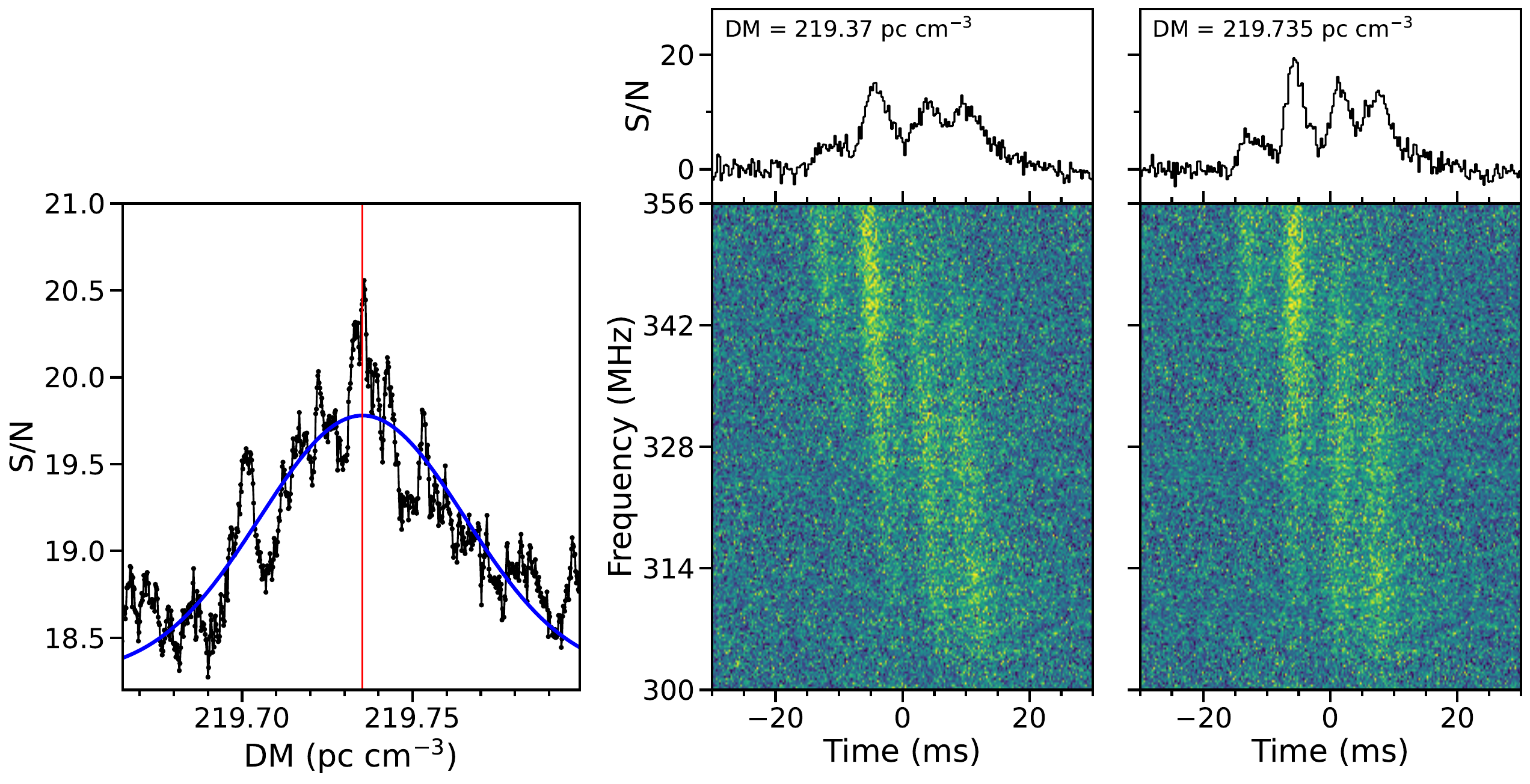}
    \caption{DM determination for bursts detected at P-band. Peak signal-to-noise (SN) versus dispersion measure (DM) curve for our brightest detection at P-band, burst B128-Wb. When applying the DM used to correct for dispersive delay for bursts detected at L-band, $219.37$~\dmunit, there still was a residual sweep present for bursts detected at P-band, as illustrated in the middle panel. By correcting the burst for range of DM values and fitting a Gaussian to the SN-curve we find a best fit DM of $219.375$~\dmunit and shown in the right panel.}
    \label{fig:dm_diff_pband}
\end{figure*}

Applying this best-fit DM on bursts detected at P-band ($330\,\mathrm{MHz}$) leaves a residual dispersive sweep, as seen in Figure~\ref{fig:dm_diff_pband}. To find the best-fit DM we dedispersed the burst to a range of trial DM values ranging from $219.665$ to $219.799$\,\dmunit and measured the peak S/N value at each trial DM. We subsequently fitted a Gaussian to the S/N-DM curve and found a best-fit DM value of $219.735$ \dmunit, as shown in Figure~\ref{fig:dm_diff_pband}.  

\subsection{Digitisation artefacts} \label{sec:digi_art} 

We record voltage data at Westerbork, Onsala and \torun in 2-bit sampling mode. The low bit depth leads to a relatively modest data rate ($\sim500 \ \textrm{GB/h}$ for $\Delta \nu \sim128$\,MHz), which allows us to observe at a high cadence and process the data with low latency. The downside of the 2-bit sampling, however, is the limited dynamic range of the samples. Digitisation artefacts can manifest in the data in case the power of a burst is concentrated in both time and frequency, for example during strong scintillation \citep{ikebe_2023_pasj, kirsten_2024_natas}. This effect can in turn result in underestimation of the energies of the bursts. We do not observe any digitisation artefacts in the 32-bit data recorded by Stockert. 

The behaviour and proposed treatment of digitisation artefacts in 2-bit sampled data has been described by \citet[][see their Section~4.1 and Fig.~4]{jenet_1998_pasp}. In essence, the 2-bit digitiser has a non-linear response to the received power, which causes the digitised signal to be underestimated compared with the total power of the un-digitised signal. This effect scales with the brightness of the signal and manifests itself as decreased power during the time of the bursts, especially during bright scintillation. This decreased power is visible as  `depressions' in the dynamic spectrum and `dips' in the frequency-averaged time series before and after the burst. In our observing setup, we quantise our data per subband, typically $8-32$\,MHz, with the result that the digitisation artefacts do not span or affect the entire observing bandwidth but are limited per subband, as visible in the top three subbands for burst B68-Wb in Figure~\ref{fig:family_plot_subset}. 

One way to correct for the digitisation artefacts is by using a dynamic level-setting scheme. This method will counter the `depression' area around the signal, but a byproduct of this method is an increase of power around the signal, which is known as quantisation noise. When applying the dynamic level-setting this increase in power is scattered uniformly across the bandwidth (in our case across a subband). Therefore, the power of the signal will be artificially increased and overestimated. This effect is best visible in the time series of a burst where power is increased before and after the burst; see Extended Data Figure~4 in \citet{kirsten_2024_natas}.  

To best estimate the power of a signal we therefore need to remove the scattered quantisation noise. This is done using a scattered power correction (SPC) algorithm, which has been implemented in the \texttt{DSPSR} software package and is applied on a per-subband basis for our recording setup. 

We create three different data products, following the same methodology we employed in \citet{kirsten_2024_natas}. For Method~\rom{1}, we use the Super FX Correlator (SFXC). We coherently (within each frequency channel) and incoherently (between frequency channels) correct the data for dispersion \citep{keimpema_2015_exa}. SFXC does not apply a dynamic level-setting scheme to the data, which means that the `dips' and `depressions' are still present in the data. Subsequently, for Method~\rom{2} we process the bursts using \texttt{digifil}, which does apply dynamic level-setting and introduces quantisation noise. Lastly, for Method~\rom{3} we apply the SPC algorithm to the data products made using \texttt{digifil} in order to compensate for the quantisation noise. 

\begin{figure*}
    \centering
    \includegraphics[width=\textwidth]{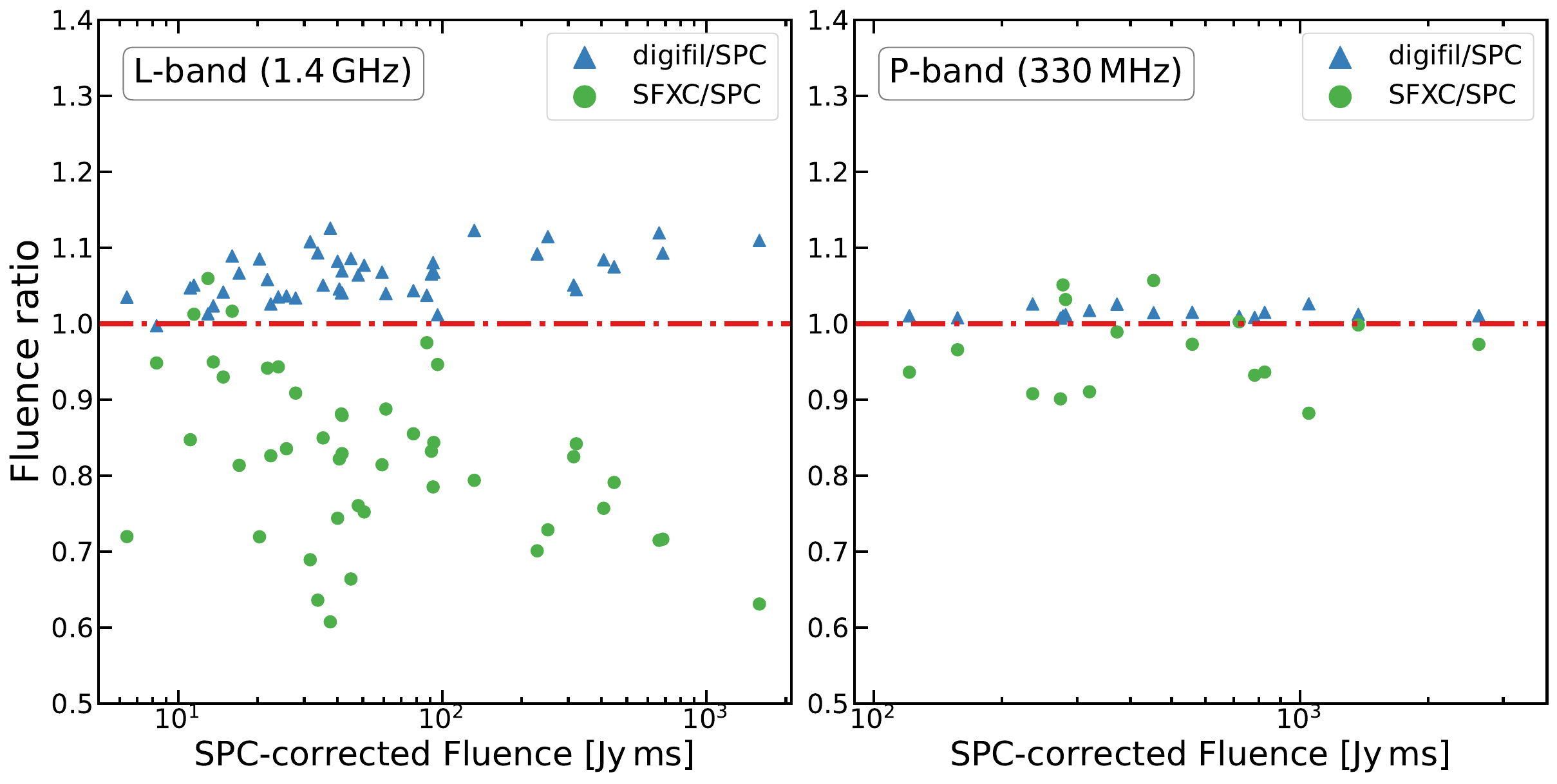}
    \caption{Ratio of fluences computed from SFXC- and digifil-generated filterbanks. The fluence for each burst was measured three different ways due to digitisation artefacts present in the data. These filterbanks were made for bursts detected with Westerbork and \torun using \texttt{SFXC}, \texttt{digifil} and \texttt{digifil} with a scattered power correction (SPC) applied. Here we plot the ratio of these measured fluences for L-band (left panel) and P-band (right panel). For L-band we find that due to digitisation effects the fluence of the burst can be underestimated for up to $40\%$~(\texttt{SFXC}) and overestimated for $10\%$~(\texttt{digifil}). Applying the SPC algorithm to correct for digitisation effects is therefore essential. Even though we detect bright ($>1000$~Jy~ms) bursts at P-band, these digitisation effects are less apparent, this most likely due to longer dispersive sweep of the burst and the low sensitivity of the Westerbork P-band receiver.}
    \label{fig:fluence_ratio_band}
\end{figure*}

In Figure \ref{fig:fluence_ratio_band}, we show the ratio of fluences relative to those measured using Method~\rom{3} for bursts detected at L-band and P-band. For L-band, we find that applying Method~\rom{1} underestimates the fluences by up to $40\%$, while Method~\rom{2} overestimates them by up to $10\%$. For P-band, these digitisation effects are less apparent which is most likely due to the longer dispersive sweep of the burst and low sensitivity of the P-band receiver at Wb. 

\bsp	
\label{lastpage}
\end{document}